\documentclass[aps,prd,floatfix,showpacs,showkeys,superscriptadress,amsmath,unsortedaddress,nofootinbib]{revtex4-1}
\pdfoutput=1
\usepackage{amssymb,bbold,subfigure}
\usepackage{amsmath,mathtools}
\usepackage{slashed}
\usepackage{color,bm}
\usepackage[dvipsnames]{xcolor}
\usepackage{tensor}
\usepackage{float}
\usepackage{titlesec}
\usepackage{hyperref}
\hypersetup{
    colorlinks=true,
    linkcolor=NavyBlue,
    filecolor=MidnightBlue,      
    urlcolor=RoyalBlue,
    citecolor=Blue,
    bookmarks=true,
    pdfpagemode=UseThumbs,
    bookmarks=true,
    pdfmenubar=true,
    pdftoolbar=true,	
}

\allowdisplaybreaks
\def\be {\begin{equation}}
\def\ee {\end{equation}}
\def\bea {\begin{eqnarray}}
\def\eea {\end{eqnarray}}
\def\beal {\begin{eqnarray}}
\def\eeal {\end{eqnarray}}
\def\bc {\begin{center}}
\def\ec {\end{center}}
\def\({\left(}
\def\){\right)}
\def\[{\left[}
\def\]{\right]}
\def\nn {\nonumber\\}
\newcommand{\sumint}[1]{{\sum\hspace{-5mm}\underset{{#1}}{\int}\hspace{2mm}}}

\newcommand{\p}{\prime}
\newcommand{\sss}[1]{{\scriptscriptstyle #1}}

\newcommand{\vp}{\mathbf{p}}
\newcommand{\om}{\omega}
\newcommand{\nf}{n_{\scriptscriptstyle{F}}}
\newcommand{\olp}{\omega_{\scriptscriptstyle{L(+)}}}
\newcommand{\olm}{\omega_{\scriptscriptstyle{L(-)}}}
\newcommand{\orp}{\omega_{\scriptscriptstyle{R(+)}}}
\newcommand{\orm}{\omega_{\scriptscriptstyle{R(-)}}}
\newcommand{\qlp}{q_{\scriptscriptstyle{L(+)}}}
\newcommand{\qlm}{q_{\scriptscriptstyle{L(-)}}}
\newcommand{\qrp}{q_{\scriptscriptstyle{R(+)}}}
\newcommand{\qrm}{q_{\scriptscriptstyle{R(-)}}}
\newcommand{\zlp}[1]{Z_{\scriptscriptstyle{L(+)}}^{#1}}
\newcommand{\zrp}[1]{Z_{\scriptscriptstyle{R(+)}}^{#1}}
\newcommand{\zlm}[1]{Z_{\scriptscriptstyle{L(-)}}^{#1}}
\newcommand{\zrm}[1]{Z_{\scriptscriptstyle{R(-)}}^{#1}}
\begin{document}

\title{\textbf{Hard dilepton production from a weakly magnetized hot QCD medium}}
\author{Aritra Das,}
\email[]{aritra.das@saha.ac.in} 
\affiliation{HENPP Division, Saha Institute of Nuclear Physics, HBNI, 
	1/AF Bidhan Nagar, Kolkata 700064, India.}
\author{Najmul Haque,}
\email[]{nhaque@niser.ac.in}
\affiliation{School of Physical Sciences, National Institute of Science Education and Research,\\ HBNI, Jatni 752050, India}
	\author{Munshi G. Mustafa}
\email[]{munshigolam.mustafa@saha.ac.in}
\affiliation{Theory Division, Saha Institute of Nuclear Physics, HBNI, 
	1/AF Bidhan Nagar, Kolkata 700064, India.} 
\author{Pradip K. Roy,}
\email[]{pradipk.roy@saha.ac.in}
\affiliation{HENPP Division, Saha Institute of Nuclear Physics, HBNI, 
  1/AF Bidhan Nagar, Kolkata 700064, India.}

\date{\today}
\begin{abstract}
	We have computed the hard dilepton production rate from a weakly magnetized deconfined QCD medium within one-loop photon self-energy by considering one hard and one thermomagnetic resummed quark propagator in the loop. In the presence of the magnetic field, the resummed propagator leads to four quasiparticle modes. The production of hard dileptons consists of rates when all four quasiquarks originating from the poles of the propagator individually annihilate with a hard quark coming from a bare propagator in the loop. Besides these, there are also contributions from a mixture of pole and Landau cut part. In weak field approximation, the magnetic field appears as a perturbative correction to the thermal contribution. Since the calculation is very involved, for a first effort as well as for simplicity, we obtained the rate up to first order in the magnetic field, {\it i.e.}, ${\cal O}[(eB)]$, which causes a marginal improvement over that in the absence of magnetic field.
\end{abstract}

\maketitle

\section{Introduction} \label{intro}

Heavy ion collisions (HIC) experiments being conducted at the LHC at CERN and the Relativistic Heavy Ion Collider at Brookhaven National Laboratory have ample evidence of the production of deconfined QCD matter at extreme conditions of high temperature and density, which is commonly termed as quark qluon plasma (QGP). This short-lived deconfined state of QCD matter has been the subject of intense investigation over the past few decades. \\

In noncentral HIC, extremely strong  magnetic field of the order of QCD scale is believed to have generated due to the presence of so-called spectator particles which do not participate in the interaction~\cite{Skokov:2009qp,Greif:2017irh}. The presence of external magnetic field is responsible for a bunch of exotic phenomena like chiral magnetic effects~\cite{Kharzeev:2013ffa,Fukushima:2012vr,Basar:2012gm}, inverse magnetic catalysis~\cite{Preis:2012fh, Farias:2014eca}, 
magnetic catalysis~\cite{Gusynin:1995nb, Shovkovy:2012zn}, superconductivity in the vacuum~\cite{Chernodub:2012tf}, and many more. Also, the thermodynamic properties of a hot magnetized deconfined QCD medium has been studied~\cite{Karmakar:2018aig,Bandyopadhyay:2017cle,Karmakar:2019tdp}. At the initial stage of the high-energy HIC, the temperature ($T_0$) is of the order of ($400-600$) MeV, and the strength of the magnetic field is approximately $15 m_\pi^2 \,  \sim T_0^2 \sim (550 {\mbox{MeV}})^2$. Nevertheless, it decays very rapidly with time, e.g., a factor of $10$ roughly within $1$ fm/$c$, beyond which it remains more or less constant over a few fm$/c$. So, it becomes extremely difficult to analyze the case with an arbitrary magnetic field. For the sake of theoretical simplicity, one works in extreme limits of a strong and a weak field regime. Apart from the temperature scale associated with the heat bath, the introduction of background magnetic field invokes another scale into the system. The strong and weak field regimes are recognized by the scales $|q_{f}B|\gg T^2\gg m^2_f$ and $T^2\gg m^2_f\gg |q_{f}B|$, respectively. 
It should be noted that in the weak field approximation, the chiral condensate vanishes and $m_f$ becomes the current quark mass.\\

QGP is a many-particle system that shows collectivity and most of its evidences are circumstantial. So, the direct detection of the QGP medium is not possible mainly due to two reasons. The first one is the fact that it exists for a very short time and 
the second one is the color confinement. Thus, one needs to rely on the direct probes like electromagnetic probes, viz., photon and dilepton, and indirect hard probes like bound states of heavy quarks, jets, collective flows etc~\cite{Wong:1995jf}  to extract its properties. One of the most popular theoretical tools is n-points current-current correlations functions that can be related to the  photon and dilepton production. The thermal dileptons are considered to be an excellent probe of the QGP medium. The reason is that it interacts only electromagnetically with the medium and leaves the medium without any final-state interaction due to its large mean free path. Also it is produced throughout  the entire volume of space-time and almost all stages of HICs. But there exist various sources of these emitted dileptons during the evolution of the created fireball.The various sources are Drell-Yan processes~\cite{Drell:1970wh}, bremsstrahlung and absorption of jets by plasma~\cite{Rapp:1999us}, and the thermal production from QGP phase. The important parameter used to characterize the emitted dilepton spectrum is its invariant mass ($M$) that can be broadly divided in three distinct ranges, namely, low with $M<M_{\phi}(=1.024\,\mbox{GeV})$, intermediate with $M_{\phi}<M<M_{J/\psi}(=3.1\,\mbox{GeV})$, and high ($M>M_{J/\psi}$). The intermediate mass range is important for getting the QGP signature and in this region the radiation from QGP dominates the mass spectrum~\cite{Chatterjee:2009rs}. \\

The theoretical calculations of the production rate of dilepton in many different scenarios of high temperature and finite chemical potential~\cite{Majumder:2000jr} proceed through the imaginary part of the two-point correlation function of the  photon~\cite{McLerran:1984ay,Weldon:1990iw}. One of the earliest seminal works in the framework of hard thermal loop perturbation theory can be found in Ref.~\cite{Braaten:1990wp}. It  calculates the rate of production of soft dilepton (lepton pair with momentum scale of the order of $gT$) using the resummed one-loop quark propagators and effective vertices.  An extensive investigation has also been carried out for small invariant mass in both LO and higher order in early the literature~\cite{Ghiglieri:2014kma, Thoma:1997dk, Greiner:2010zg, Aurenche:2002pc}.  
In Ref.~\cite{Greiner:2010zg}, the low invariant mass ($M\ll\!1\,\mbox{GeV}$) thermal dilepton rates  have been investigated from the deconfined QCD phase using perturbative and nonperturbative methods.The low mass dilepton rate has also been computed in Ref.~\cite{Bandyopadhyay:2015wua}, considering both electric and magnetic scale resummation via the Gribov formalism. As noted earlier, owing to the presence of external magnetic field in noncentral heavy ion collision, there is enough motivation to investigate the behavior of electromagnetic probes under the influence of a background magnetic field~\cite{Tuchin:2013ie, Tuchin:2012mf, Tuchin:2013bda}. Recently there has been some detailed investigation of the dilepton rate from the one loop-photon polarization tensor. In Ref.~\cite{Sadooghi:2016jyf}, the production rate of the dilepton has been computed using the Ritus Eigenfunction method~\cite{Ritus:1972ky}. On the other hand Refs~\cite{Bandyopadhyay:2016fyd,Bandyopadhyay:2017raf} have investigated dilepton production in a hot magnetized medium using weak~\cite{Chyi:1999fc} and strong field approximation of the quark propagator~\cite{Gusynin:1995nb}, whereas Ref.~\cite{Ghosh:2018xhh} has calculated it using the full form of the Schwinger propagator. It has also been computed using the effective QCD model in the presence of an external magnetic field~\cite{Islam:2018sog}.

The straightforward extension to the case in which hard dileptons are considered can be found in Ref.~\cite{Turbide:2006mc}. In this calculation, it has been argued that it is sufficient to consider one resumed propagator ({\it i.e.,} soft) and one hard propagator in one-loop photon self-energy. The reason is that since the momentum flowing through the external photon line is hard, one of the quark propagators, which must have hard momentum flowing through it, can be taken as bare. But for the other propagator, one should take the resummed (i.e., soft) propagator. In this paper, we follow the same line in which we use one magnetic field-dependent free propagator and one thermomagnetic resummed propagator for obtaining the hard dilepton rate from a weakly magnetized deconfined QCD medium.\footnote{For having soft dilepton, one can use both the propagators as well as all the vertices effective, but the calculation will be extremely involved and complicated. However, as a first effort and also for simplicity, we consider one magnetic field dependent free propagator and one resummed propagator in one loop photon self energy, 
which itself is an indeed very involved calculation as we will see below.}

The paper is organized as follows. In Sec.~\ref{sec:formalism}, we briefly outline the notation used and also the quark propagator in the presence of a weak background field. The dispersion properties of a resummed quark propagator and its spectral density in the presence of a weakly magnetized hot medium are discussed in  Sec.~\ref{sec:disp}. The calculation of dilepton production  and results are given in details in Sec.~\ref{sec:dilep_prod} at zero magnetic field (in Sec.~\ref{sec:dilep_prod_zero}) and at weak magnetic field in a thermalized background (in Sec.~\ref{sec:dilep_prod_nonzero}). Finally, we conclude in Sec.~\ref{concl}.

\section{Notations and charged fermion propagator in background magnetic field within Schwinger formalism} \label{sec:formalism}
We begin by defining the following notation for the $4$-vector and the metric tensor:
\begin{align*}
  & a^{\mu}=(a^0,a^1,a^2,a^3), \qquad g^{\mu\nu}=\mbox{diag}(1,-1,-1,-1),\\
  & g^{\mu\nu}_{\sss{\parallel}}=\mbox{diag}(1,0,0,-1), \qquad g^{\mu\nu}_{\sss{\perp}}=\mbox{diag}(0,-1,-1,0),\\
  & g^{\mu\nu}=g^{\mu\nu}_{\sss{\parallel}}+g^{\mu\nu}_{\sss{\perp}},\qquad \slashed{a}=\gamma^{\mu}a_{\mu}, \\
  & \slashed{a}_{\sss{\parallel}}=\gamma^{0}a^{0}-\gamma^{3}a^{3}, \qquad \slashed{a}_{\sss{\perp}}=(\bm{\gamma\cdot a})_\perp=\gamma^{1}a^{1}+\gamma^{2}a^{2},\\
  & a^{\mu}_{\sss{\parallel}}=g^{\mu\nu}_{\sss{\parallel}}a^{\nu}, \qquad a^{\mu}_{\sss{\perp}}=-g^{\mu\nu}_{\sss{\perp}}a^{\nu}.
   \end{align*}
The Green's function satisfying the Dirac equation in the presence of a magnetic field can be written as
\begin{align}
(i\slashed{\partial}-Qq_f\slashed{A}_{ext}(x)-m_f){G}(x,x^{\prime})=\delta^{(4)}(x-x^{\prime}), \label{dirac_equation}
\end{align}
where $A_{ext}$ is the vector potential for an external background magnetic field, $Q$ is ${\mathrm {sgn}}(q_fB)$, and $q_f$ is the absolute value of the particle's charge and $m_f$ is the mass of a particle. This equation can be solved by different methods, as, for example, Schwinger's proper time method~\cite{Schwinger:1951nm}, the Ritus eigenfunction method~\cite{Ritus:1972ky}, the Furry's picture~\cite{Furry:1951zz} for the case of a constant field pointing in the $z$ direction. The Green's function in equation~\eqref{dirac_equation} can be written as
\beal
  {G}(x,x^{\prime}) = \Phi(x,x^{\prime})\int\frac{d^4K}{(2\pi)^4}\exp(-iK\cdot x)\tilde{\mathcal{G}}(K), \label{gen_green}
  \eeal
 where the prefactor $\Phi(x,x^{\prime})$ is a phase factor that breaks both translational and gauge invariance. However, it can be taken as unity by choosing the symmetric gauge of vector potential, {\it i.e.}, $\displaystyle A_{ext}^{\mu}=\frac{B}{2}\(0,y,-x,0\)$.
 So, the momentum space Green's function \textit{vis-\`a-vis} the propagator is written as~\cite{Gusynin:1995nb,Chyi:1999fc}
  \bea
  \tilde{\mathcal{G}}(K) = \exp\left(-\frac{k^2_{\sss{\perp}}}{q_{f}B}\right)\sum^{\infty}_{n=0}\frac{D_{n}(q_fB,K)}{k^2_0-2n q_fB-k_z^2-m_f^2}, \label{miransky_full}
  \eea
  where
  \beal
  D_n(q_fB,K)&\equiv \left(\slashed{K}_{\sss{\parallel}}+m_f\right)\Bigg[\Big(1-i\mbox{sgn}(q_fB)\gamma^1\gamma^2\Big)
  L_n\left(\frac{k^2_{\sss{\perp}}}{q_fB}\right)-\Big(1+i\mbox{sgn}(q_fB)\gamma^1\gamma^2\Big)L_{n-1}\left(\frac{k^2_{\sss{\perp}}}{q_fB}\right)\Bigg]\nonumber \\
  &+4\slashed{k}_{\sss{\perp}}L^1_{n-1}\left(2\frac{k^2_{\sss{\perp}}}{|q_fB|}\right),
  \eeal
 where $L_n^\alpha(x)$ is the generalized Laguerre polynomial.
 As stated earlier, we are interested in the domain $T^2\gg m_f^2 \gg q_fB$. 
 In this domain, one can approximate the propagator by expanding the sum over $n$ in 
 Eq.~\eqref{miransky_full} in the power of $q_fB$ to obtain a simplified form~\cite{Chyi:1999fc}  as
  \beal
  S_{\sss{F}}(K)&=&\frac{\slashed{K}+m_f}{K^2-m_f^2}+i\gamma^1\gamma^2\frac{\slashed{K}_{\sss{\parallel}}+m_f}{(K^2-m_f^2)^2}\,q_fB
  +\mathcal{O}[(q_fB)^2] \nonumber \\
 &=& S_F^{(0)}(K)+S_F^{(1)}(K) + {\mathcal O}[(q_fB)^2],\label{weak_bare_propagator}
  \eeal
  where $S_F^{(0)}$ is the ${\cal O}[(q_fB)^0]$ and $S_F^{(1)}$ is the ${\cal O}[(q_fB)]$ part of the propagator $S_F$.
 \subsection{Dispersion of fermionic modes and spectral representation} \label{sec:disp}
The dispersion behavior of the resummed fermionic propagator in the presence of weak magnetic field is discussed in our earlier work~\cite{Das:2017vfh}. Here in this section, we shall briefly outline some important results that
would be useful here for the sake of clarity. The effective propagator is given by 
\bea
S^*(K)=\mathcal{P}_-\frac{\slashed{L}(K)}{L(K)^2}\mathcal{P}_+ + \mathcal{P}_+\frac{\slashed{R}(K)}{R(K)^2}\mathcal{P}_- \label{weak_eff_prop} ,
\eea
with $4$-momentum $K\equiv(k_0,{\mathbf k})= \(k_0,k_{\sss{\perp}},k_z\)$ with $|{\mathbf k}|\equiv k=\sqrt{k_{\sss{\perp}}^2+k_z^2}$ and where the chirality projection operators are given as
\bea
\mathcal{P}_\pm = \frac{1}{2}\(\mathbb{1}\pm \gamma_5\).
\eea
$\slashed{L}$ and $\slashed{R}$ that appear in Eq.~\eqref{weak_eff_prop} can be written in the rest frame of the heat bath along with the magnetic field in the $z$ direction as 
\bea
\slashed{L} &=& \[(1+a(k_0,k))k_0+b(k_0,k)+b'(k_0,k_\perp,k_z)\]\gamma^0 - \left[(1+a(k_0,k))k_z+c'(k_0,k_\perp,k_z)\right]\gamma^3\nn
&& - (1+a(k_0,k))(\bm{\gamma\cdot k})_\perp \nn
&=& \left[(1+a(k_0,k))k_0+b(k_0,k)+b'(k_0,k_\sss{\perp},k_z)\right]\gamma^0 - \left[k(1+a(k_0,k))\right](\bm{\gamma\cdot\hat{k}})- c'(k_0,k_\perp,k_z)\gamma^3\nn
&=& g_L^1(k_0,k_\perp,k_z)\gamma^0 - g_L^2(k_0,k_\sss{\perp},k_z)(\bm{\gamma \cdot \hat{k}}) - g_L^3(k_0,k_\sss{\perp},k_z)\gamma^3, \label{l_rest} \\
\slashed{R} &=& \left[(1+a(k_0,k))k_0+b(k_0,k)-b'(k_0,k_\sss{\perp},k_z)\right]\gamma^0 - \left[(1+a(k_0,k))k_z-c'(k_0,k_\sss{\perp},k_z)\right]\gamma^3 \, \nn
&& -(1+a(p_0,p))(\bm{\gamma \cdot k})_\perp \nn
&=& \left[(1+a(k_0,k))k_0+b(k_0,k)-b'(k_0,k_\perp,k_z)\right]\gamma^0 - \left[k(1+a(k_0,k))\right](\bm{\gamma \cdot \hat{k}})+ c'(k_0,k_\perp,k_z)\gamma^3 \nn
&=& g_R^1(k_0,k_\sss{\perp},k_z)\gamma^0 - g_R^2(k_0,k_\perp,k_z)(\bm{\gamma\cdot \hat{k}}) + g_R^3(k_0,k_\sss{\perp},k_z)\gamma^3 \, , \label{r_rest}
\eea
where ${\hat {\bm k}} ={\mathbf k}/k$. 
Although, $g_L^2=g_R^2;~g_L^3=g_R^3$, for the sake of convenience, they are treated separately as $g_L^i$ and $g_R^i$.

The pole of the effective fermion propagator $S^{*}(K)$ in weak magnetized media gives the dispersion relation of fermionic mode. 
The dispersion equations are given by
  \begin{align}
    L^2=L_{+}L_{-}=0, \qquad R^2=R_{+}R_{-}=0 , \nonumber
  \end{align}
  where $L_{\pm}$ and $R_{\pm}$ are, respectively, given by
  \begin{align}
    L_{\pm}\(k_0,k_\perp,k_z\)&=(1+a)k_0+b+b^{\prime}\mp\[\(1+a)k_z+c^{\prime}\)^2+(1+a)^2k^{2}_{\sss{\perp}}\]^{1/2}, \\
    R_{\pm}\(k_0,k_\perp,k_z\)&=(1+a)k_0+b-b^{\prime}\mp\[\(1+a)k_z-c^{\prime}\)^2+(1+a)^2k^{2}_{\sss{\perp}}\]^{1/2}.
  \end{align}
The forms of the structure functions~\cite{Das:2017vfh} are quoted here as
   \begin{align}
a& =  -\frac{m^{2}_{th}}{k^2}Q_{1}\(\frac{k_0}{k}\), \label{at}\\ 
b& =  \frac{m^{2}_{th}}{k}\[\frac{k_{0}}{k}Q_{1}\(\frac{k_{0}}{k}\)-Q_{0}\(\frac{k_{0}}{k}\)\], \label{bt} \\
b' & =  4C_{F}g^{2}M^{2}(T,m_f,q_{f}B)\frac{k_{z}}{k^{2}}Q_{1}\(\frac{k_{0}}{k}\), \label{bprime}\\
c' &= 4C_{F}g^{2}M^{2}\!\(T,m_f,q_{f}B\)\frac{1}{k}Q_{0}\left(\frac{k_{0}}{k}\right), \label{cprime}
\end{align}
where the $Q$'s can  be found in Ref.~\cite{Das:2017vfh},
the thermomagnetic mass is given as~\cite{Haque:2017nxq,Ayala:2014uua}
\begin{align}
M^{2}(T,m_f,q_{f}B) &= \frac{q_{f}B}{16\pi^{2}}\[\ln(2)-\frac{T}{m_f}\frac{\pi}{2}\] \, , \label{magneticmass}
\end{align}
and also the thermal mass is given as
\begin{align}
  m_{th}^{2}=\frac{1}{8}C_{F}g^{2}T^{2}. 
  \end{align}
\begin{figure}[H]
\centering
\includegraphics[width=0.45\textwidth]{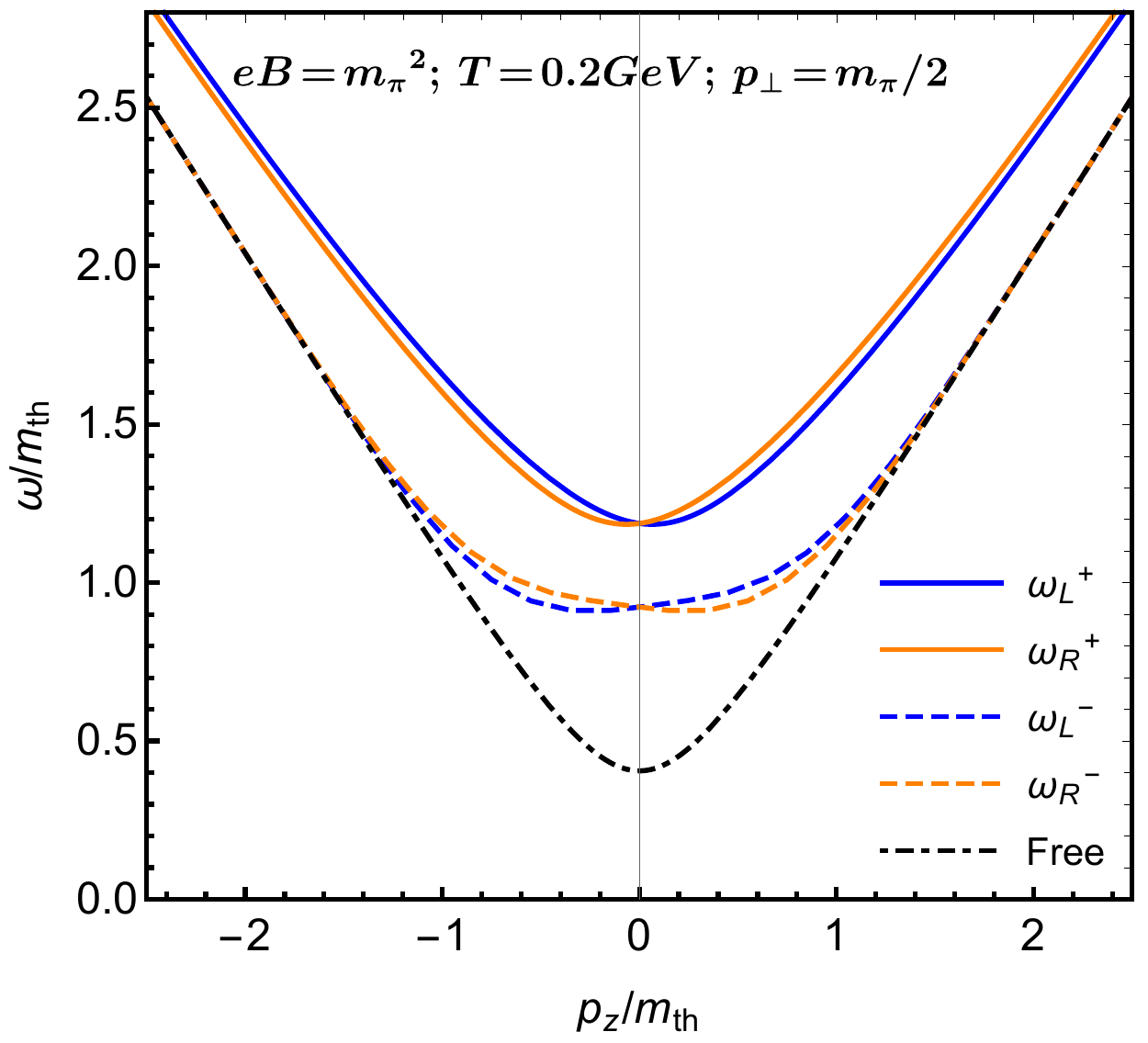}
\includegraphics[width=0.45\textwidth]{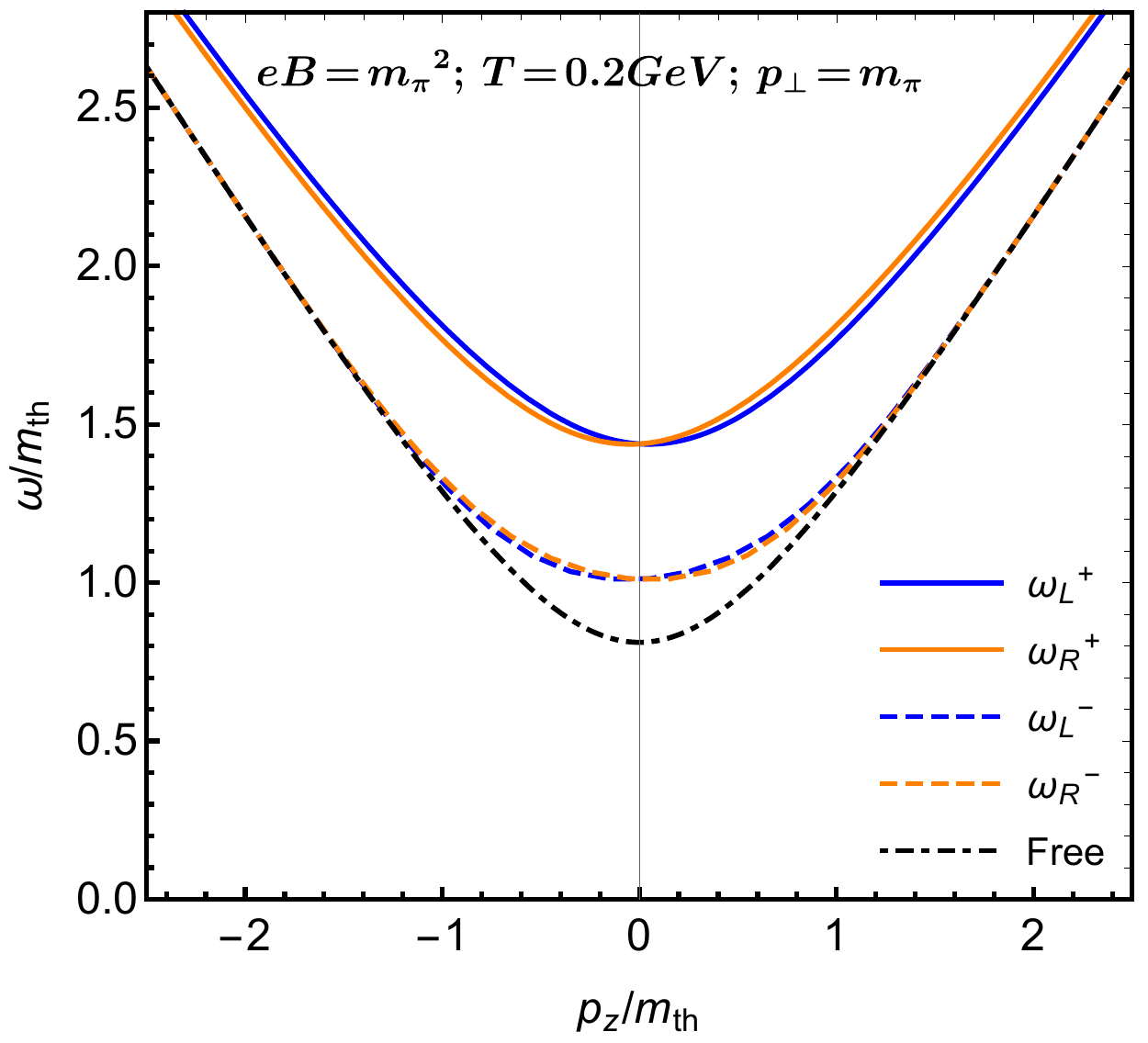}
\caption{ This displays the various $u$-quark dispersion modes. The free dispersion of hard quark $q$ with energy 
$\omega =\sqrt{p_z^2+p_\perp^2}$ with $p_\perp=m_\pi/2$ (left panel) and $m_\pi$ (right panel).}
\label{fig:HLLfig}
\end{figure}
The dispersion solutions~\cite{Das:2017vfh} are noted as a function of $p_{\sss{\perp}}$ and $p_{z}$ as
\begin{align}
  L_{+} &= 0 \Longrightarrow p_{0}=\Big(\olp, -\orm\Big), \label{lp}\\
  L_{-} &= 0 \Longrightarrow p_{0}=\Big(\olm, -\orp\Big), \label{lm}\\
  R_{+} &= 0 \Longrightarrow p_{0}=\Big(\orp, -\olm\Big), \label{rp}\\
  R_{-} &= 0 \Longrightarrow p_{0}=\Big(\orm, -\olp\Big). \label{rm}
\end{align}
The corresponding dispersion of various quark modes $\qlp$, $\qlm$, $\qrp$ and $\qrm$ with respective frequencies $\olp$, $\olm$, $\orp$ and $\orm$ are displayed in Fig.~\ref{fig:HLLfig}. The free dispersion of hard quark $q$ with energy 
$\omega =\sqrt{p_z^2+p_\perp^2}$ is also displayed. It is clear from Fig.~\ref{fig:HLLfig} that the processes that we expect will involve one hard and one soft quark since we are using one free (hard) quark  propagator in the presence
of magnetic field and one resummed thermomagnetic quark  (soft) propagator in Fig.~\ref{fig:dilepton_production}. Now, one can write the various dilepton production processes from the dispersion plot as $q \qlp \longrightarrow \gamma^*\longrightarrow l^+l^-$, $q \qlm \longrightarrow \gamma^*\longrightarrow l^+l^-$, 
$q \qrp \longrightarrow \gamma^*\longrightarrow l^+l^-$, and $q \qrm \longrightarrow \gamma^*\longrightarrow l^+l^-$. There could also be soft decay processes like $\qlp \longrightarrow q \gamma^*\longrightarrow q l^+l^-$, $\qlm \longrightarrow q \gamma^*\longrightarrow q l^+l^-$,
$\qrp \longrightarrow q \gamma^*\longrightarrow q l^+l^-$, and $\qrm \longrightarrow q \gamma^*\longrightarrow q l^+l^-$. We will see below that all of them may not be allowed due to kinematical restrictions. Also, besides these processes
there will be soft processes from Landau cut contributions. We will discuss these contributions in detail later.

\subsection{Spectral function of quark propagator}
For computation of the dilepton rate, the spectral function of the quark propagator is needed. The spectral representation of the effective quark propagator in a hot magnetized medium is obtained in Ref.~\cite{Das:2017vfh}. We briefly outline both the quark propagator and it's spectral representation here.

Now, the effective propagator in Eq.~\eqref{weak_eff_prop} can be decomposed into six parts by separating 
out the $\gamma$ matrices as
\bea
S^*(k_0,k_{\sss\perp},k_z) &=& \mathcal{P}_-\gamma^0\mathcal{P}_+~\frac{g_L^1(k_0,k_{\sss\perp},k_z)}{L^2} 
- \mathcal{P}_-(\bm{\gamma\cdot \hat{k}})\mathcal{P}_+~\frac{g_L^2(k_0,k_\sss\perp,k_z)}{L^2} 
- \mathcal{P}_-\gamma^3\mathcal{P}_+~\frac{g_L^3(k_0,k_\sss\perp,k_z)}{L^2} \nn
&+& \mathcal{P}_+\gamma^0\mathcal{P}_-~\frac{g_R^1(k_0,k_\sss\perp,k_z)}{R^2} 
- \mathcal{P}_+(\bm{\gamma\cdot \hat{k}})\mathcal{P}_-~\frac{g_R^2(k_0,k_\sss\perp,k_z)}{R^2} 
+ \mathcal{P}_+\gamma^3\mathcal{P}_-~\frac{g_R^3(k_0,k_\sss\perp,k_z)}{R^2}.
\label{prop_pre_spec}
\eea
It was discussed earlier that $L^2=0$ yields four poles, giving four modes with positive and negative energy, $\omega_{L^{(\pm)}}(k_\sss\perp,k_z)$ and $-\omega_{R^{(\pm)}}(k_\sss\perp,k_z)$, as given in Eqs.~\eqref{lp} and~\eqref{lm}.Similarly, $R^2=0$  also gives four poles, namely $\omega_{R^{(\pm)}}(k_\sss\perp,k_z)$ and $-\omega_{L^{(\pm)}}(k_\sss\perp,k_z)$, as given in Eqs.~\eqref{rp} and~\eqref{rm}. With this information, the spectral 
representation~\cite{Bellac:2011kqa,Das:2017vfh,Karsch:2000gi,Chakraborty:2001kx,Braaten:1990wp} is obtained for the effective propagator in Eq.~\eqref{prop_pre_spec} as
\bea
\rho &=& \left(\mathcal{P}_-\gamma^0\mathcal{P}_+\right)~\rho_L^1 
- \left( \mathcal{P}_-(\bm{\gamma\cdot\hat{k}})\mathcal{P}_+\right)~\rho_L^2 
- \left( \mathcal{P}_-\gamma^3\mathcal{P}_+\right)~\rho_L^3 \nn
&& + \left( \mathcal{P}_+\gamma^0\mathcal{P}_-\right)~\rho_R^1
 - \left(\mathcal{P}_+(\bm{\gamma\cdot\hat{k}})\mathcal{P}_-\right)~\rho_R^2 
 + \left( \mathcal{P}_+\gamma^3\mathcal{P}_-\right)~\rho_R^3 \, ,
\label{prop_spec}
\eea
where the spectral functions corresponding to each of the terms can be written as 
\bea
\rho_L^i &=& \frac{1}{\pi} ~\mathrm{Im}\left(\frac{g_L^i}{L^2}\right) = \frac{1}{\pi} ~\mathrm{Im}\left(F_{L}^{i}\right)\nn
 &=& \zlp{i+}(k_\sss\perp,k_z)\delta(k_0-\olp(k_\sss\perp,k_z))+\zlm{i+}(k_\sss\perp,k_z)\delta(k_0-\olm(k_\sss\perp,k_z))\nn
 &&+\zrm{i-}(k_\sss\perp,k_z)\delta(k_0+\orm(k_\sss\perp,k_z))+\zrp{i-}(k_\sss\perp,k_z)\delta(k_0+\orp(k_\sss\perp,k_z))+\beta_L^i
 \, ,\label{spec_li}\\
 \rho_R^i &=& \frac{1}{\pi} ~\mathrm{Im}\left(\frac{g_R^i}{R^2}\right) = \frac{1}{\pi} ~\mathrm{Im}\left(F_{R}^{i}\right)\nn
  &=& \zrp{i+}(k_\sss\perp,k_z)\delta(k_0-\orp(k_\perp,k_z))+\zrm{i+}(k_\sss\perp,k_z)\delta(k_0-\orm(k_\sss\perp,k_z))\nn
 &&+\zlm{i-}(k_\sss\perp,k_z)\delta(k_0+\olm(k_\sss\perp,k_z))+\zlp{i-}(k_\sss\perp,k_z)\delta(k_0+\olp(k_\sss\perp,k_z))+\beta_R^i\, ,
 \label{spec_ri}
 \eea
 where $i=1,2,3$. The delta functions are originated from the timelike domain ($k_0^2 >k^2)$ whereas the cut parts $\beta^i_{L(R)}$ are involved with the Landau damping originating from the spacelike domain ($k_0^ 2 < k^2$) of the propagator. The residues $Z^i_{L(R)}$ are determined at the various poles as
 \bea
 Z_{L(R)}^{i \ {\mbox{sgn of pole }}}(k_\sss\perp,k_z) = g_{L(R)}^i(k_0,k_\sss\perp,k_z) \Bigg| \frac{\partial L^2(R^2)}{\partial k_0} \Bigg|^{-1}_{k_0=\mbox{ pole}} \, ,\label{residue}
 \eea
where the expressions of residues can be written~\cite{Das:2017vfh} in terms of the structure coefficients $a$, $b$, $b^{\p}$, and $c^{\p}$ and their derivatives.
 \section{Dilepton production} 
 \label{sec:dilep_prod}

 \begin{figure}[h!]
	\centering
	\includegraphics[scale=0.4]{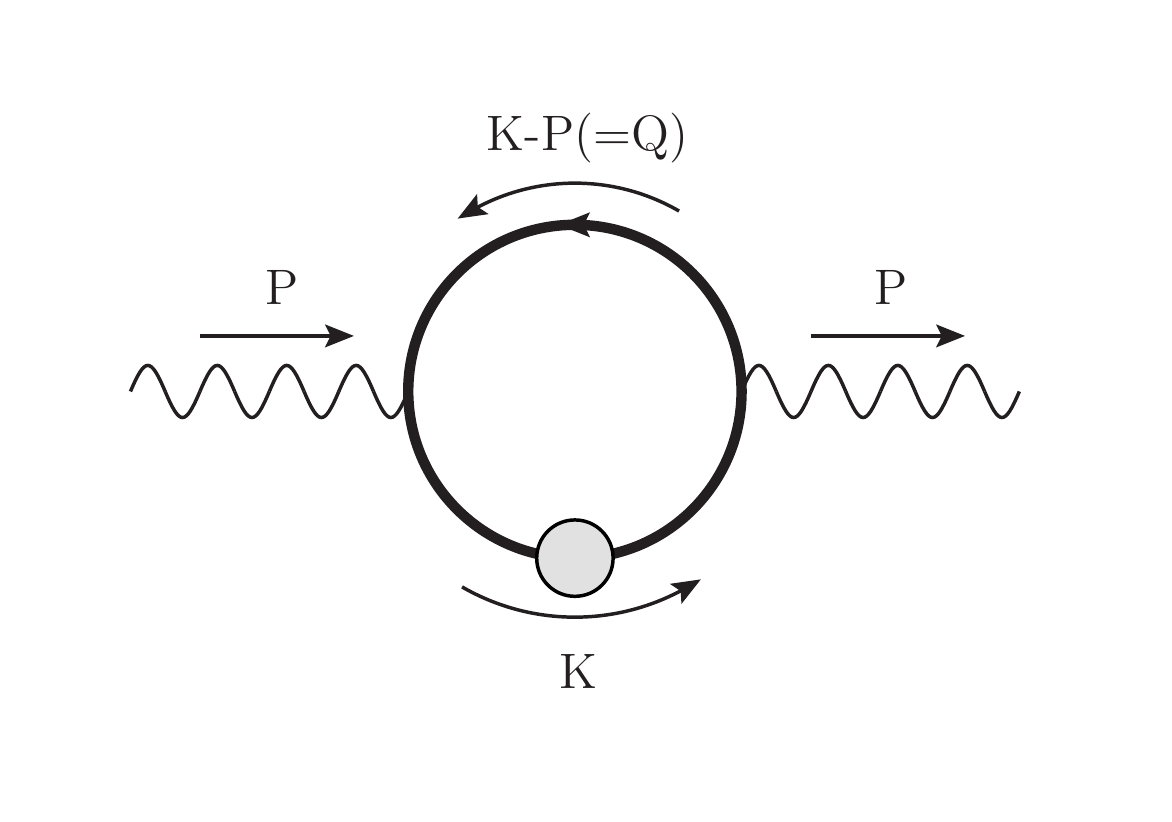}
	\caption{Feynman diagram for the production of the hard dileption in presence of weak background magnetic field}
	\label{fig:dilepton_production}
 \end{figure}
 The differential dilepton production can be written as~\cite{Braaten:1990wp,Greiner:2010zg}
\begin{align}
\frac{dR}{d^4xd^4P}=\frac{\alpha}{12\pi^4}\frac{1}{P^2}\frac{1}{e^{\beta p_0}-1}\mbox{Im}\Pi\indices{^\mu_\mu}(p_0+i\epsilon,p), \label{dilepfor}
\end{align}
with $P^2\equiv p_0^2-p^2=M^2$ where $M$ is the invariant mass of the dilepton.
Now, for simplification we will consider the case with $p=0$.

The expression for one-loop self-energy can be obtained from the Feynman diagram in the Fig.~\ref{fig:dilepton_production} as
\begin{align}
\Pi\indices{^\mu^\nu}(P)=-N_ce^2\sum_f\left(\frac{q_f}{e}\right)^2\sumint{K}\mbox{Tr}\left[\gamma\indices{^\mu}S_F(Q)\gamma\indices{^\nu}S^{*}(K)\right],
\label{pimn}
\end{align} 
where $N_c=3$ is color factor and $Q\equiv K-P$. In imaginary time formalism the loop integral  can be written as
\bea
\int \frac{d^4K}{(2\pi)^4} &\equiv&  \sumint{K} = T\sum_{k_0}\int\frac{d^3k}{(2\pi)^3}.
\eea


\subsection{Dilepton rate at vanishing magnetic field} 
\label{sec:dilep_prod_zero}
\begin{figure}[tbh]
	\includegraphics[scale=0.6]{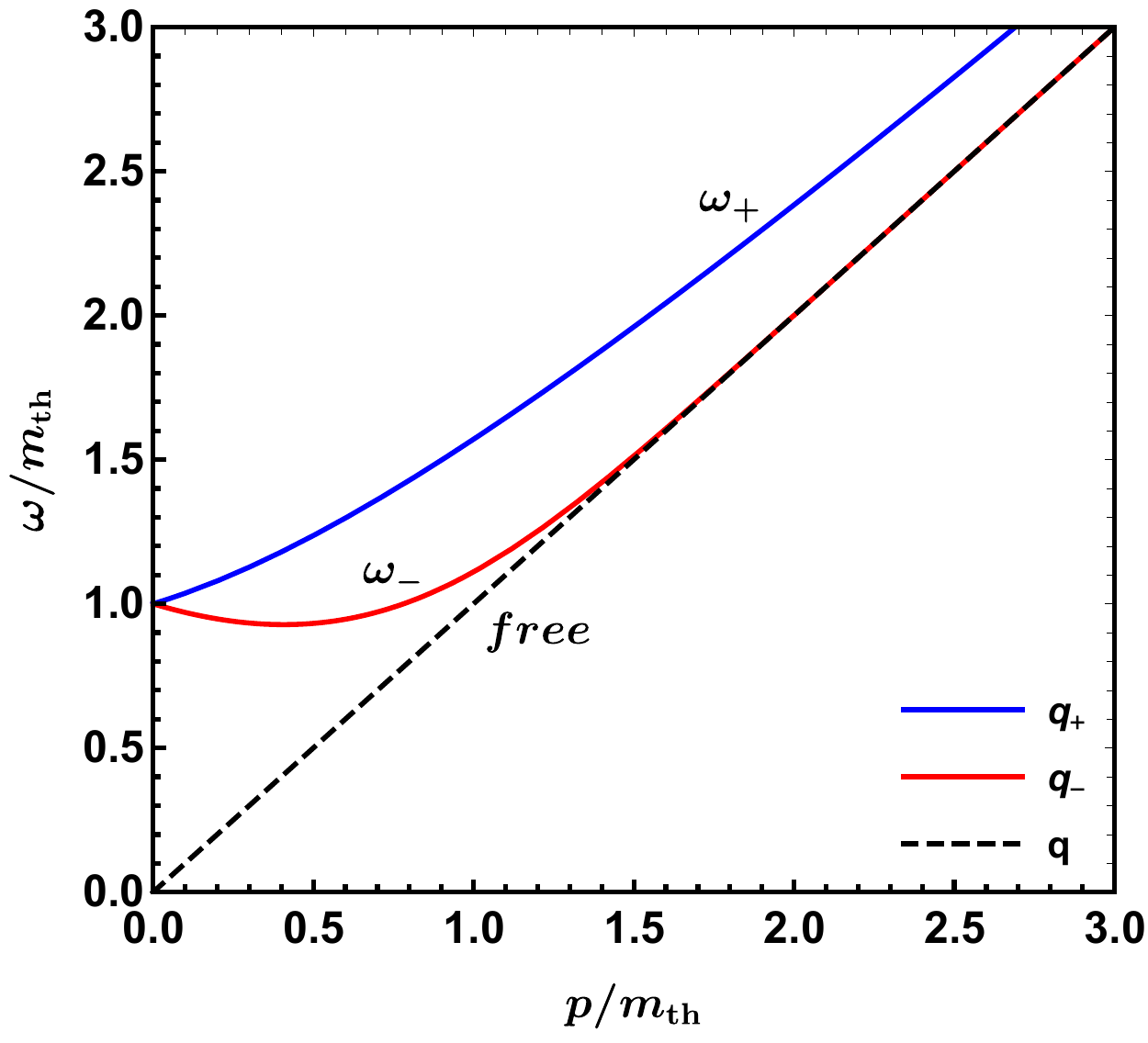}
	\caption{Soft (HTL) and hard (free) quark dispersion relation. $q_+$ and $q_-$ are soft quarks coming from HTL resummed propagator and $q$ is hard quark coming from free propagator.}
	\label{fig:htl_dispersion}
\end{figure}

In this section, we first discuss the dilepton production rate without any external magnetic field. For this purpose, we use one hard quark propagator and Hard Thermal Loop (HTL) resummed soft quark propagator with two modes~\cite{Bellac:2011kqa} : one quasiquark mode $q_+$ with energy $\omega_+$ and other a plasmino mode $q_-$ with energy $\omega_-$. The free hard quark is represented by $q$ with energy $k$. The corresponding dispersion is shown in Fig.~\ref{fig:htl_dispersion}. Now, in this case the allowed dilepton production processes coming from pole-pole part are annihilation processes $qq_+\longrightarrow \gamma^*\longrightarrow l^+l^-$ and soft decay process $q_-\longrightarrow q \gamma^*\longrightarrow q l^+l^-$. There will also be other processes which are not allowed by  energy conservation and kinematical restriction with the photon momentum, $\bm p=0$.
In addition, there will also be pole-cut contributions, as will be discussed below in detail. We also note that there is no cut-cut contribution as the spectral function for the hard propagator has only pole contributions. Now, the one-loop photon self-energy $\Pi^\mu_\mu$ with one hard propagator $S_0$ and one resummed HTL propagator $S_{\text{HTL}}$ can be written as
\begin{eqnarray}
\Pi^\mu_\mu&=&-N_ce^2\sum_f\(\frac{q_f}{e}\)^2\sumint{K}\mbox{Tr}\left[\gamma\indices{^\mu}S_0(K)\gamma\indices{_\mu}S_{\text{HTL}}(Q)\right]\nonumber\\
&=&2N_ce^2T\sum_f\(\frac{q_f}{e}\)^2\sum_{k_0}\int\frac{d^3k}{(2\pi)^3}\left[\frac{1}{D_+(k)}\left(\frac{1-\bm{\hat{k}\cdot\hat{q}}}{d_+(q)}
+\frac{1+\bm{\hat{k}\cdot\hat{q}}}{d_-(q)}\right)+\frac{1}{D_-(k)}\left(\frac{1+\bm{\hat{k}\cdot\hat{q}}}{d_+(q)}
+\frac{1-\bm{\hat{k}\cdot\hat{q}}}{d_-(q)}\right)\right],\label{Pimm_eB0}
\end{eqnarray} 
with
\bea
d_\pm(q_0,q)&=&q_0-q\\
D_\pm(k_0,k)&=& k_0\mp k -\frac{m_{\rm th}^2}{2k}\left[\(1\mp\frac{k_0}{k}\)\log\frac{k_0+k}{k_0-k}  \pm 2\right].
\eea
Now, the imaginary part of Eq.~(\ref{Pimm_eB0}) is obtained as
\bea
\rm{Im}\Pi_\mu^\mu&=& 2N_ce^2T\sum_f\(\frac{q_f}{e}\)^2\left(e^{E/T}-1\right)\nn
&\times& \int\frac{d^3k}{(2\pi)^3}\int_{-\infty}^{\infty}d\omega \int_{-\infty}^{\infty}d\omega'\delta(E-\omega-\omega')
n_F(\omega)n_F(\omega')\pi\left[(1-\bm{\hat{k}\cdot\hat{q}})(\rho_+r_-+\rho_-r_+)\right.\nn
&&\hspace{8cm}+\left.(1+\bm{\hat{k}\cdot\hat{q}})(\rho_+r_++\rho_-r_-)\right],
\eea
which at $\bm{p}=0$ reads as 
\bea
\rm{Im}\Pi_\mu^\mu&=& 2N_ce^2T\pi\sum_f\(\frac{q_f}{e}\)^2\left(e^{E/T}-1\right)\nn
&\times& \int\frac{d^3k}{(2\pi)^3}\int_{-\infty}^{\infty}d\omega \int_{-\infty}^{\infty}d\omega'\delta(E-\omega-\omega')n_F(\omega)n_F(\omega')2(\rho_+r_++\rho_-r_-).
\eea
The spectral representations of soft and hard propagator read~\cite{Bellac:2011kqa}, respectively, as
\bea
\rho_\pm(\omega,k)&=& \frac{\omega^2-k^2}{2m_{th}^2}\left[\delta(\omega-\omega_\pm(k))+\delta(\omega+\omega_\mp(k))\right] + \beta_\pm(\omega,k)\Theta(k^2-\omega^2),\\
r_\pm(\omega',k)&=&\delta(\omega'\mp k),
\eea
with
\bea
\beta_\pm(x,y)= \frac{1}{2}\frac{y\mp x}{ \left[y (x\mp y)-\frac{1}{2} \left(1\mp \frac{x}{y}\right) \log \left| \frac{x+y}{x-y}\right| \mp 1\right]^2+\left[\frac{1}{2} \pi  \left(1\mp \frac{x}{y}\right)\right]^2},
\eea
where $x=\omega/m_{th}$ and $y=k/m_{th}$.
The soft spectral function contains the pole part coming from the poles of the HTL propagator and Landau cut contribution from the spacelike domain, $k^2<\omega^2$, of the HTL propagator. The hard spectral function has only pole parts. So, there will be  four energy conserving $\delta$ functions from the pole-pole part, namely, 
$\delta(E+\omega_++k)$, $\delta(E-\omega_-+k)$, $\delta(E-\omega_++k)$ and $\delta(E-\omega_+-k)$. But two  processes $q q_+ \gamma^*\longrightarrow  \text{nothing}$ and $q\longrightarrow q_- \gamma^*\longrightarrow q_- l^+l^-$ 
coming, respectively, from $\delta(E+\omega_++k)$ and $\delta(E+\omega_--k)$ are not allowed by the energy conservation. The remaining two allowed processes coming from
$\delta(E-\omega_+-k)$ and $\delta(E-\omega_-+k)$ lead to the respective processes $qq_+\longrightarrow \gamma^*\longrightarrow l^+l^-$ and $q_-\longrightarrow q \gamma^*\longrightarrow q l^+l^-$ as discussed earlier. The resulting pole-pole part of the dilepton rate is
\bea
\left.\frac{dR}{d^4xd^4P}\right|_{\rm{pole-pole}}&=&\frac{\alpha}{12\pi^4}\frac{1}{E^2}\frac{1}{e^{\beta E}-1} 12\pi e^2\sum_f\(\frac{q_f}{e}\)^2\left(e^{E/T}-1\right) \int\frac{d^3k}{(2\pi)^3}\nn
&\times&\left[\frac{\omega_+^2-k^2}{2m_{\rm{th}}^2}n_F(\omega_+)n_F(k)\delta(E-\omega_+-k)+\frac{\omega_-^2-k^2}{2m_{\rm{th}}^2}n_F(\omega_-)n_F(-k)\delta(E-\omega_-+k)\right]\nn
&=& \frac{2\alpha^2}{\pi^4E^2}\sum_f\(\frac{q_f}{e}\)^2\int k^2dk\nn
&\times&\left[\frac{\omega_+^2-k^2}{2m_{\rm{th}}^2}n_F(\omega_+)n_F(k)\delta(E-\omega_+-k)+\frac{\omega_-^2-k^2}{2m_{\rm{th}}^2}n_F(\omega_-)n_F(-k)\delta(E-\omega_-+k)\right].
\eea
Scaling $\omega_{\pm}, k$ with $m_{th}$ as $x_\pm=\omega_{\pm}/m_{th}$, $E_s=E/m_{th}$ and  we get
\bea
\left.\frac{dR}{d^4xd^4P}\right|_{\rm{pole-pole}}&=&  \frac{\alpha^2}{\pi^4E_s^2}\sum_f\(\frac{q_f}{e}\)^2\int y^2dy\Bigg[\left(x_+^2-y^2\right)\frac{1}{e^{\beta m_{th} x_+}+1}\frac{1}{e^{\beta m_{th} y}+1}\delta\left(E_s-x_+-y\right)\nn
&&+\left(x_-^2-y^2\right)\frac{1}{e^{\beta m_{th} x_-}+1}\frac{1}{e^{-\beta m_{th} y}+1}\delta\left(E_s-x_-+y\right)\Bigg].
\eea
Now, the pole-cut part of the rate is obtained as
\bea
\left.\frac{dR}{d^4xd^4P}\right|_{\rm{pole-cut}}&=&\frac{\alpha}{12\pi^4}\frac{1}{E^2}\frac{1}{e^{\beta E}-1} 12\pi e^2\sum_f\(\frac{q_f}{e}\)^2\left(e^{E/T}-1\right) \int\frac{d^3k}{(2\pi)^3}\int_{-k}^{k}d\omega\nn
&\times&\left[\beta_+(\omega,k)n_F(\omega)n_F(k)\delta(E-\omega-k)+\beta_-(\omega,k)n_F(\omega)n_F(-k)\delta(E-\omega+k)\right]\nn
&=&  \frac{2\alpha^2}{\pi^4E_s^2}\sum_f\(\frac{q_f}{e}\)^2\int y^2dy\int_{-y}^{y}dx\nn
&\times&\left[\beta_+(x,y)n_F(x)n_F(y)\delta(E_s-x-y)+\beta_-(x,y)n_F(x)n_F(-y)\delta(E_s-x+y)\right].
\eea
We note that the second term of the pole-cut rate will vanish as the delta function gives the condition $x=E_s+y$, which lies outside of the domain $-y\le x\le y$ and the pole-cut contribution becomes
\bea
\left.\frac{dR}{d^4xd^4P}\right|_{\rm{pole-cut}}
=  \frac{2\alpha^2}{\pi^4E_s^2}\sum_f\(\frac{q_f}{e}\)^2\int y^2dy\  
\beta_+(E_s-y,y)n_F(E_s-y)n_F(y)\Theta(2y-E_s).
\eea
\begin{figure}[tbh]
	\includegraphics[scale=0.7]{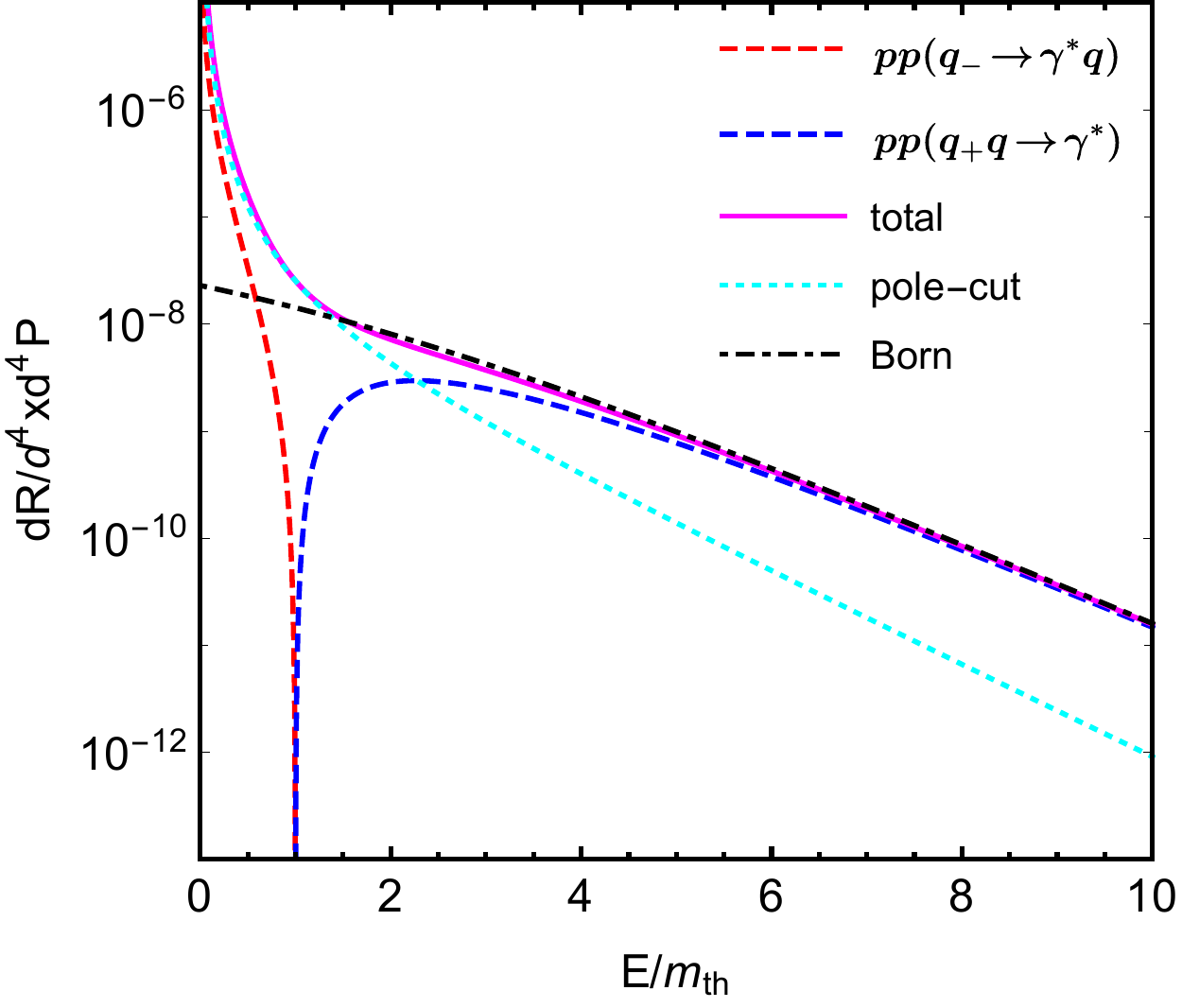}
	\caption{Dilepton rate for vanishing magnetic field}
	\label{fig:dilepton_rate_eb0}
\end{figure}
It is worth it to write the Born rate~\cite{Greiner:2010zg}  as
\bea
\left.\frac{dR}{d^4xd^4P}\right|_{\rm born} = \sum_f\(\frac{q_f}{e}\)^2\frac{\alpha^2}{4\pi^4}n_F^2(E/2).
\eea
In Fig.~\ref{fig:dilepton_rate_eb0}, we display the dilepton rate in the absence of magnetic field. For  $E=0$ the dilepton rate begins with the transition process $q_-\longrightarrow q \gamma^*\longrightarrow q l^+l^-$. This rate begins with a divergence as all plasmino, $q_-$, modes with higher energy (Fig.~\ref{fig:htl_dispersion}) prefer to make the transition to a free quark mode with lower energy and thus the density of states diverges. However, this rate decays very first because the plasmino mode $q_-$ is exponentially suppressed and merges with the free hard quark mode as shown in Fig.~\ref{fig:htl_dispersion}. Then the annihilation of one soft ($q_+$) and one hard ($q$) mode, $qq_+\longrightarrow \gamma^*\longrightarrow l^+l^-$, begins when $E=m_{th}$ (as the mass of the hard mode is zero). It then grows with $E$ and matches with the Bonn rate at large $E$. The dilepton rate coming from pole-cut part dominates at low $E$ and falls off below  the Bonn rate at large $E$. The net rate dominates the Bonn rate at low energy.

\subsection{Dilepton rate at finite magnetic field} \label{sec:dilep_prod_nonzero}
In this section, we shall investigate dilepton production in the presence of weak homogeneous background magnetic field. We are concerned about the dilepton whose momenta are of the order of $T$, i.e., $p_0, p \sim T$. In that case, as discussed, we need to dress just one quark propagator~\cite{Turbide:2006mc} as in Fig.~\ref{fig:dilepton_production}. The bare propagator in the weak magnetic field approximation is given in Eq.~\eqref{weak_bare_propagator}. The dressed propagator is given in Eq.~\eqref{weak_eff_prop}, which, for convenience, is decomposed into two parts as
\be
S^*(K)=S^{*}_{L}(K)+S^{*}_{R}(K),  \label{s_eff_decom}
\ee
where 
\begin{align}
S^{*}_{L}(K)=\mathcal{P}_{-}\frac{\slashed{L}}{L^2}\mathcal{P}_{+}, \qquad S^{*}_{R}(K)=\mathcal{P}_{+}\frac{\slashed{R}}{R^2}\mathcal{P}_{-}.
\end{align}
Now, using Eqs.~\eqref{weak_bare_propagator} and~\eqref{s_eff_decom},
the one-loop photon polarization tensor in Eq.~\eqref{pimn} corresponding to Fig.~\ref{fig:dilepton_production} can be obtained as
\begin{align}
\Pi\indices{^{\mu}_{\mu}}(p_0,\vp) &= - N_ce^2\sum_f \left(\frac{q_f}{e}\right)^2 \sumint{K} \mbox{Tr}\big[\gamma^{\mu}S^{*}(K)\gamma_{\mu}S_{F}(Q)\big] \nn
&=-N_ce^2\sum_f \left(\frac{q_f}{e}\right)^2 \sumint{K} \mbox{Tr}\big[\gamma^{\mu}S^{*}_{L}(K)\gamma_{\mu}S^{(0)}_{F}(Q)\big]-N_ce^2
\sum_f \left(\frac{q_f}{e}\right)^2 \sumint{K} \mbox{Tr}\big[\gamma^{\mu}S^{*}_{R}(K)\gamma_{\mu}S^{(0)}_{F}(Q)\big] \nn
&-N_ce^2\sum_f \left(\frac{q_f}{e}\right)^2 \sumint{K} \mbox{Tr}\big[\gamma^{\mu}S^{*}_{L}(K)\gamma_{\mu}S^{(1)}_{F}(Q)\big]-N_ce^2
\sum_f \left(\frac{q_f}{e}\right)^2 \sumint{K} \mbox{Tr}\big[\gamma^{\mu}S^{*}_{R}(K)\gamma_{\mu}S^{(1)}_{F}(Q)\big].
\label{Pimm_eB}
\end{align}
The result of the Dirac trace is 
\begin{align}
\mbox{Tr}\big[\gamma^{\mu}S^{*}(K)\gamma_{\mu}S_{F}(Q)\big] &=  -4 \Bigg[\frac{L^{\mu}Q_{\mu}}{L^2(Q^2-m_{\sss{f}}^2)}+\frac{R^{\mu}Q_{\mu}}{R^2(Q^2-m_{\sss{f}}^2)}+q_{f}B\Bigg\lbrace \frac{Q^0L^3-Q^3L^0}{L^2(Q^2-m_{\sss{f}}^2)^2}-\frac{Q^0R^3-Q^3R^0}{R^2(Q^2-m_{\sss{f}}^2)^2}\Bigg\rbrace\Bigg],
\end{align} 
where $m_{\sss{f}}$ is the current quark mass. The components of $L^{\mu} = (L^{0},L^{1},L^{2},L^{3})$ and $R^{\mu} = (R^{0},R^{1},R^{2},R^{3})$ are given by 
\begin{align}
L^{0} &= \[1+a(k_0,k)\]k_0+b(k_0,k)+b^{\p}(k_0,k_{\sss{\perp}},k_z), \nn
L^{i} &= \[1+a(k_0,k)\]k^i; \qquad\qquad	 i=1,2 \nn
L^{3} &= \[1+a(k_0,k)\]k_z + c^{\p}(k_0,k), \nn
R^{0} &= \[1+a(k_0,k)\]k_0+b(k_0,k)-b^{\p}(k_0,k_{\sss{\perp}},k_z), \nn
R^{i} &= (1+a(k_0,k))k^i; \qquad\qquad i=1,2 \nn
R^{3} &= (1+a(k_0,k))k_z - c^{\p}(k_0,k). \label{li_ri}
\end{align}
Now Eq.~\eqref{li_ri} can be expressed in terms of $g_{L,R}^{i}$ ($i=1,2,3$) as
\begin{align}
L^{0} &= g_{L}^{1}(k_0,k_{\sss{\perp}},k_z), \nn
L^{i} &= g_{L}^{2}(k_0,k)\hat{k}^{i}; \qquad\qquad	 i=1,2 \nn
L^{3} &= g_{L}^{2}(k_0,k)\hat{k}^{3}+g_{L}^{3}(k_0,k), \nn
R^{0} &= g_{R}^{1}(k_0,k_{\sss{\perp}},k_z), \nn
R^{i} &= g_{R}^{2}(k_0,k)\hat{k}^{i}; \qquad\qquad i=1,2 \nn
R^{3} &= g_{R}^{2}(k_0,k)\hat{k}^{3}-g_{R}^{3}(k_0,k).
\end{align} 
As discussed in the previous subsection, we will investigate the case in which the virtual photon is at rest in the plasma rest frame, 
i.e., $\bm{p}=\mathbf{0}$, $P^{\mu}=(p_0,\mathbf{0})$. In this case, $Q^{\mu}=K^{\mu}-P^{\mu}=(k_0-p_0,\bm{k})$. Thus, Eq.~\eqref{Pimm_eB} becomes
\begin{align}
\Pi\indices{^{\mu}_{\mu}}(p_0,\mathbf{0})&=12e^2\sum_f \left(\frac{q_f}{e}\right)^2 \sumint{K} \Bigg[\frac{L_0(k_0-p_0)-\bm{L\cdot k}}{L^2[(k_0-p_0)^2-\omega_{k}^{2}]}+\frac{R_0(k_0-p_0)-\bm{R \cdot k}}{R^2[(k_0-p_0)^2-\omega_{k}^{2}]}\nn
&\hspace{3cm}+q_{\sss{f}}B\Bigg\lbrace\frac{L_z(k_0-p_0)-k_zL_0}{L^2[(k_0-p_0)^2-\omega_{k}^{2}]^2}-\frac{R_z(k_0-p_0)-k_zR_0}{R^2[(k_0-p_0)^2-\omega_{k}^{2}]^2}\Bigg\rbrace\Bigg] \nn
&=12e^2\sum_f \left(\frac{q_f}{e}\right)^2\sumint{K} \Bigg[\frac{(k_0-p_0)g_{L}^1-k g_{L}^{2}-k_zg_{L}^{3}}{L^2[(k_0-p_0)^2-\omega_{k}^{2}]}+\frac{(k_0-p_0)g_{R}^1-k g_{R}^{2}+k_zg_{R}^{3}}{R^2[(k_0-p_0)^2-\omega_{k}^{2}]} \nn
&\hspace{3cm}+q_{\sss{f}}B\frac{k_zg_{L}^1-(k_0-p_0)(\hat{k}_zg_L^2+g_L^3)}{L^2[(k_0-p_0)^2-\omega_{k}^{2}]^2} - q_{\sss{f}}B\frac{k_zg_{R}^1-(k_0-p_0)(\hat{k}_zg_R^2-g_R^3)}{R^2[(k_0-p_0)^2-\omega_{k}^{2}]^2}\Bigg] \nn
&=12e^2\sum_f \left(\frac{q_f}{e}\right)^2\sumint{K} \Bigg[\frac{k_0-p_0}{(k_0-p_0)^2-\omega_{k}^2}F_{L}^{1}-k \frac{1}{(k_0-p_0)^2-\omega_{k}^2}F_{L}^{2}-k_z\frac{1}{(k_0-p_0)^2-\omega_{k}^2}F_{L}^{3} \nn
&\hspace{1cm}+\frac{k_0-p_0}{(k_0-p_0)^2-\omega_{k}^2}F_{R}^{1}
-k \frac{1}{(k_0-p_0)^2-\omega_{k}^2}F_{R}^{2}+k_z\frac{1}{(k_0-p_0)^2-\omega_{k}^2}F_{R}^{3}\nn
&\hspace{0.2cm}+q_{\sss{f}}B\Bigg\lbrace k_z \frac{1}{[(k_0-p_0)^2-\omega_{k}^2]^2}F_{L}^1-\hat{k}_z \frac{k_0-p_0}{[(k_0-p_0)^2-\omega_{k}^2]^2}F_{L}^2 -\frac{k_0-p_0}{[(k_0-p_0)^2-\omega_{k}^2]^2}F_{L}^3\nn
&\hspace{1cm}-k_z\frac{1}{[(k_0-p_0)^2-\omega_{k}^2]^2}F_{R}^1+\hat{k}_{z}\frac{k_0-p_0}{[(k_0-p_0)^2-\omega_{k}^2]^2}F_{R}^2-\frac{k_0-p_0}{[(k_0-p_0)^2-\omega_{k}^2]^2}F_{R}^3\Bigg\rbrace\Bigg] \nn
&=12e^2\sum_f \left(\frac{q_f}{e}\right)^2\sumint{K} \Bigg[ f_0^{(1)}F_L^1-k f_0^{(0)}F_L^2-k_zf_0^{(0)}F_L^3+f_0^{(1)}F_R^1-k f_0^{(0)}F_R^2+k_zf_0^{(0)}F_R^3 \nn
&\hspace{1cm}+q_{\sss{f}}B\left\lbrace k_zf_1^{(0)}F_L^1-\hat{k}_zf_1^{(1)}F_L^2-f_1^{(1)}F_L^3-k_zf_1^{(0)}F_R^1+\hat{k}_zf_1^{(1)}F_R^2-f_1^{(1)}F_R^3\right\rbrace\Bigg] \nn
&=12e^2\sum_f \left(\frac{q_f}{e}\right)^2\int\frac{d^3k}{(2\pi)^3}\Bigg[T\sum_{k_0}(F_L^1+F_R^1)f_1^{(0)}-k T\sum_{k_0}(F_L^2+F_R^2)f_0^{(0)}-k_z T\sum_{k_0}(F_L^3-F_R^3)f_0^{(0)}    \nn
&\hspace{0.5in}+q_{\sss{f}}B\Big\lbrace k_z T\sum_{k_0}(F_L^1-F_R^1)f_0^{(1)}-\hat{k}_z T\sum_{k_0}(F_L^2-F_R^2)f_1^{(1)}-T\sum_{k_0}(F_L^3+F_R^3)f_1^{(1)}\Big\rbrace\Bigg]. 
\label{pimumu}
\end{align}
Here in Eq.~\eqref{pimumu}, $\omega_k\equiv\sqrt{k^2+m^2_{\sss{f}}}$ and we used the shorthand notation as
$F_{(L,R)}^{i}\equiv F_{(L,R)}^{i}(k_0,k_{\sss{\perp}},k_z)$ and $f_{0,1}^{(0),(1)} \equiv f_{0,1}^{(0),(1)} \big(k_0-p_0,k\big)$. 
Written explicitly they are given as
\begin{align}
F_L^i \equiv \frac{g_L^i}{L^2}, \qquad F_R^i \equiv \frac{g_R^i}{R^2}; \qquad\qquad i=1,2,3 \nn
f^{(0)}_{0}(k_0-p_0,k) \equiv \frac{1}{(k_0-p_0)^2-\omega_k^2}, \qquad f^{(1)}_{0}(k_0-p_0,k) \equiv \frac{k_0-p_0}{(k_0-p_0)^2-\omega_k^2}, \nn
f^{(0)}_{1}(k_0-p_0,k) \equiv \frac{1}{[(k_0-p_0)^2-\omega_k^2]^2}, \qquad f^{(1)}_{1}(k_0-p_0,k) \equiv \frac{k_0-p_0}{[(k_0-p_0)^2-\omega_k^2]^2}.
\end{align}
We take the imaginary part of Eq.~\eqref{pimumu}  with a decomposition as
\begin{align}
\mbox{Im}\Pi\indices{^{\mu}_{\mu}}(p'_0,\mathbf{0}) = \mbox{Im}\Pi\indices{^{1\mu}_{\mu}}(p'_0,\mathbf{0})-\mbox{Im}\Pi\indices{^{2\mu}_{\mu}}(p'_0,\mathbf{0})-\mbox{Im}\Pi\indices{^{3\mu}_{\mu}}(p'_0,\mathbf{0})+\mbox{Im}\Pi\indices{^{4\mu}_{\mu}}(p'_0,\mathbf{0})-\mbox{Im}\Pi\indices{^{5\mu}_{\mu}}(p'_0,\mathbf{0})-\mbox{Im}\Pi\indices{^{6\mu}_{\mu}}(p_0,\mathbf{0}), \label{pimumu_together} 
\end{align}
where $p'_0=p_0+i\epsilon$. The various terms on the rhs of the above equation are defined as
\begin{align}
\mbox{Im}\Pi\indices{^{1\mu}_{\mu}}(p'_0,\mathbf{0}) &= 12e^2\sum_f \left(\frac{q_f}{e}\right)^2\int\frac{d^3k}{(2\pi)^3} \mbox{Im}\,T\sum_{k_0}\left[F_{L}^{1}(k_0,k_{\sss{\perp}},k_z)+F_{R}^{1}(k_0,k_{\sss{\perp}},k_z)\right]f^{(1)}_{0}(k_0-p'_0,k), \label{pimumu1}\\
\mbox{Im}\Pi\indices{^{2\mu}_{\mu}}(p'_0,\mathbf{0}) &= 12e^2\sum_f \left(\frac{q_f}{e}\right)^2\int\frac{d^3k}{(2\pi)^3} k\mbox{Im}\,T\sum_{k_0}\left[F_{L}^{2}(k_0,k_{\sss{\perp}},k_z)+F_{R}^{2}(k_0,k_{\sss{\perp}},k_z)\right]f^{(0)}_{0}(k_0-p'_0,k), \label{pimumu2}\\
\mbox{Im}\Pi\indices{^{3\mu}_{\mu}}(p'_0,\mathbf{0}) &= 12e^2\sum_f \left(\frac{q_f}{e}\right)^2\int\frac{d^3k}{(2\pi)^3} k_z\mbox{Im}\,T\sum_{k_0}\left[F_{L}^{3}(k_0,k_{\sss{\perp}},k_z)-F_{R}^{3}(k_0,k_{\sss{\perp}},k_z)\right]f^{(0)}_{0}(k_0-p'_0,k), \label{pimumu3}\\
\mbox{Im}\Pi\indices{^{4\mu}_{\mu}}(p'_0,\mathbf{0}) &= 12e^2\sum_f \left(\frac{q_f}{e}\right)^2q_{\sss{f}}B\int\frac{d^3k}{(2\pi)^3} k_z\mbox{Im}\,T\sum_{k_0}\left[F_{L}^{1}(k_0,k_{\sss{\perp}},k_z)-F_{R}^{1}(k_0,k_{\sss{\perp}},k_z)\right]f^{(0)}_{1}(k_0-p'_0,k), \label{pimumu4}\\
\mbox{Im}\Pi\indices{^{5\mu}_{\mu}}(p'_0,\mathbf{0}) &= 12e^2\sum_f \left(\frac{q_f}{e}\right)^2q_{\sss{f}}B\int\frac{d^3k}{(2\pi)^3} \hat{k}_z\mbox{Im}\,T\sum_{k_0}\left[F_{L}^{2}(k_0,k_{\sss{\perp}},k_z)-F_{R}^{2}(k_0,k_{\sss{\perp}},k_z)\right]f^{(1)}_{1}(k_0-p'_0,k), \label{pimumu5}\\
\mbox{Im}\Pi\indices{^{6\mu}_{\mu}}(p'_0,\mathbf{0}) &= 12e^2\sum_f \left(\frac{q_f}{e}\right)^2q_{\sss{f}}B\int\frac{d^3k}{(2\pi)^3} \mbox{Im}\,T\sum_{k_0}\left[F_{L}^{3}(k_0,k_{\sss{\perp}},k_z)+F_{R}^{3}(k_0,k_{\sss{\perp}},k_z)\right]f^{(1)}_{1}(k_0-p'_0,k). \label{pimumu6}
\end{align}
Now, by applying the Braaten-Pisarski-Yuan prescription~\cite{Braaten:1990wp}, the imaginary parts of Eqs.~\eqref{pimumu1} -\eqref{pimumu6} can be obtained in 
terms of the spectral function of the propagators [Eqs.~\eqref{spec_li},~\eqref{spec_ri},~\eqref{f_0^1},~\eqref{f_0^0},~\eqref{f_1^1} and \eqref{f_1^0}] as
\begin{align}
\mbox{Im}\Pi\indices{^{1\mu}_{\mu}}(p'_0,\mathbf{0}) &= 12e^2\sum_f \left(\frac{q_f}{e}\right)^2\pi\left(1-e^{\beta p_0}\right)\int\frac{d^3k}{(2\pi)^3}\int_{-\infty}^{\infty} d\om \int_{-\infty}^{\infty}d\om^{\p}\left[\rho_{L}^{1}(\om)+\rho_{R}^{1}(\om)\right]\rho_{0}^{(1)}(-\om^{\p})\nn
&\hspace{8cm}\times n_{\sss{F}}(\om)n_{\sss{F}}(\om^{\p})\delta(p_0-\om-\om^{\p}), \label{spec_pimumu1} \\
\mbox{Im}\Pi\indices{^{2\mu}_{\mu}}(p'_0,\mathbf{0}) &= 12e^2\sum_f \left(\frac{q_f}{e}\right)^2\pi\left(1-e^{\beta p_0}\right)\int\frac{d^3k}{(2\pi)^3}k \int_{-\infty}^{\infty} d\om \int_{-\infty}^{\infty} d\om^{\p}\left[\rho_{L}^{2}(\om)+\rho_{R}^{2}(\om)\right]\rho_{0}^{(0)}(-\om^{\p})\nn 
&\hspace{8cm}\times n_{\sss{F}}(\om)n_{\sss{F}}(\om^{\p})\delta(p_0-\om-\om^{\p}),  \label{spec_pimumu2} \\
\mbox{Im}\Pi\indices{^{3\mu}_{\mu}}(p'_0,\mathbf{0}) &= 12e^2\sum_f \left(\frac{q_f}{e}\right)^2\pi\left(1-e^{\beta p_0}\right)\int\frac{d^3k}{(2\pi)^3}k_z\int_{-\infty}^{\infty} d\om \int_{-\infty}^{\infty}d\om^{\p}\left[\rho_{L}^{3}(\om)-\rho_{R}^{3}(\om)\right]\rho_{0}^{(0)}(-\om^{\p}) \nn 
&\hspace{8cm}
\times n_{\sss{F}}(\om)n_{\sss{F}}(\om^{\p})\delta(p_0-\om-\om^{\p}), \label{spec_pimumu3} \\
\mbox{Im}\Pi\indices{^{4\mu}_{\mu}}(p'_0,\mathbf{0}) &= 12e^2\sum_f \left(\frac{q_f}{e}\right)^2q_{\sss{f}}B\pi\left(1-e^{\beta p_0}\right)\int\frac{d^3k}{(2\pi)^3}k_z\int_{-\infty}^{\infty} d\om \int_{-\infty}^{\infty}d\om^{\p}\left[\rho_{L}^{1}(\om)-\rho_{R}^{1}(\om)\right]\rho_{1}^{(0)}(-\om^{\p})\nn
&\hspace{8cm}\times
n_{\sss{F}}(\om)n_{\sss{F}}(\om^{\p})\delta(p_0-\om-\om^{\p}),  \label{spec_pimumu4}\\
\mbox{Im}\Pi\indices{^{5\mu}_{\mu}}(p'_0,\mathbf{0}) &= 12e^2\sum_f \left(\frac{q_f}{e}\right)^2q_{\sss{f}}B\pi\left(1-e^{\beta p_0}\right)\int\frac{d^3k}{(2\pi)^3}\hat{k}_z\int_{-\infty}^{\infty} d\om \int_{-\infty}^{\infty} d\om^{\p}\left[\rho_{L}^{2}(\om)-\rho_{R}^{2}(\om)\right]\rho_{1}^{(1)}(-\om^{\p})\nn 
&\hspace{8cm}\times n_{\sss{F}}(\om)n_{\sss{F}}(\om^{\p})\delta(p_0-\om-\om^{\p}), \label{spec_pimumu5} \\
\mbox{Im}\Pi\indices{^{5\mu}_{\mu}}(p'_0,\mathbf{0}) &= 12e^2\sum_f \left(\frac{q_f}{e}\right)^2q_{\sss{f}}B\pi\left(1-e^{\beta p_0}\right)\int\frac{d^3k}{(2\pi)^3}\int_{-\infty}^{\infty} d\om \int_{-\infty}^{\infty} d\om^{\p}\left[\rho_{L}^{3}(\om)+\rho_{R}^{3}(\om)\right]\rho_{1}^{(1)}(-\om^{\p})\nn
&\hspace{8cm}\times
n_{\sss{F}}(\om)n_{\sss{F}}(\om^{\p})\delta(p_0-\om-\om^{\p}) . \label{spec_pimumu6}
\end{align}  
As before, the rate has a pole-pole and a pole-cut part. There will also be no cut-cut  part since the spectral function for a hard quark has only the pole part. 
Below, we compute various contributions.

\subsubsection{Pole-pole part}
Here, to compute the pole-pole contribution of the dilepton rate, we divide it by two parts. The contribution coming from the free part of $S_F$ and $S^*$ is termed 
as (a) magnetic field-independent part, whereas  that coming from the $\mathcal{O}[(q_{\sss{f}}B)]$ part of $S_F$ and $S^*$ is termed as  (b) magnetic field-dependent part. Note that we neglect current quark mass $m_f$ so that $\omega_k=k$.

\vspace*{0.1in}

\noindent{\sf{(a) Magnetic field independent part:}} \\

Using Eq.~\eqref{rho^1_0} in~\eqref{spec_pimumu1}, we get,
\begin{align}
\mathrm{Im}\Pi\indices{^{1\mu}_{\mu}} &= 12e^2\sum_{f}\(\dfrac{q_{\sss{f}}}{e}\)^2\pi(1-e^{\beta p_0}) \frac{1}{(2\pi)^3}\int_{0}^{2\pi}d\phi \int_{0}^{\infty}dk\,k^2 \int_{-1}^{1}d\xi\int_{-\infty}^{\infty}d\om d\om^{\p}\left[\rho^{1}_{L}(\om)+\rho^{1}_{R}(\om)\right] \nn
&\hspace{4cm}\times\left[\frac{-\delta(\om^{\p}+k)-\delta(\om^{\p}-k)}{2}\right]n_{\sss{F}}(\om)n_{\sss{F}}(\om^{\p})\delta(p_0-\om-\om^{\p}) \nn
&=\frac{3e^2}{2\pi}(e^{\beta p_0}-1)\sum_{f}\left(\dfrac{q_{\sss{f}}}{e}\right)^2\int_{0}^{\infty}dk\,k^2 \int_{-1}^{1}d\xi \int_{-\infty}^{\infty}d\om\left[\rho^{1}_{L}(\om)+\rho^{1}_{R}(\om)\right]n_{\sss{F}}(\om) \nn
&\hspace{4cm}\times\left[\nf(-k)\delta(p_0-\om+k)+\nf(k)\delta(p_0-\om-k)\right]. \label{ptc}
\end{align}
Now, the spectral functions $\rho_{L}^{1}$ and $\rho_{R}^{1}$ have a pole part as well as a cut part. But here we will only use the pole part of the spectral functions. In the pole part, there are four terms in $\rho_{L(R)}^{1}$ [Eqs.~\eqref{spec_li} and \eqref{spec_ri}] out of which the terms with a positive sign of the pole will survive from energy conservation and we now write them as
\begin{align}
\left.\mathrm{Im}\Pi\indices{^{1\mu}_{\mu}}\right|_{\text{pole-pole}} &=\frac{3e^2}{2\pi}(e^{\beta p_0}-1)\sum_{f}\left(\dfrac{q_{\sss{f}}}{e}\right)^2\int_{0}^{\infty}dk\,k^2 \int_{-1}^{1}d\xi\int_{-\infty}^{\infty}d\om\nf(\om)\Big[Z^{1+}_{L(+)}\delta(\om-\om_{L(+)})+Z^{1+}_{L(-)}\delta(\om-\om_{L(-)}) \nn
&\hspace{2cm}+Z^{1+}_{R(+)}\delta(\om-\om_{R(+)})+Z^{1+}_{R(-)}\delta(\om-\om_{R(-)})\Big]\left[\nf(-k)\delta(p_0-\om+k)+\nf(k)\delta(p_0-\om-k)\right] \nn
&=\frac{3e^2}{2\pi}(e^{\beta p_0}-1)\sum_{f}\left(\dfrac{q_{\sss{f}}}{e}\right)^2\int_{0}^{\infty}dk\,k^2 \int_{-1}^{1}d\xi \Big[\nf(\om_{L(+)})\nf(k)\delta\(p_0-\om_{L(+)}-k\)+\nf(\om_{L(-)})\nf(k)   \nn
&\times\delta(p_0-\om_{L(-)}-k)+\nf(\om_{R(+)})\nf(k)\delta(p_0-\om_{R(+)}-k)+\nf(\om_{R(-)})\nf(k)\delta(p_0-\om_{R(-)}-k)   \nn
&+\nf(\om_{L(+)})\nf(-k)\delta(p_0-\om_{L(+)}+k)+\nf(\om_{L(-)})\nf(-k)\delta(p_0-\om_{L(-)}+k) \nn
&+\nf(\om_{R(+)})\nf(-k)\delta(p_0-\om_{R(+)}+k) +\nf(\om_{R(-)})\nf(-k)\delta(p_0-\om_{R(-)}+k)\big].
\end{align}
Now, in a similar manner and using Eq.~\eqref{rho^0_0} in Eq.~\eqref{spec_pimumu2}, we get
\begin{align}
\left.\mathrm{Im}\indices{^{2\mu}_{\mu}}  \right|_{\text{pole-pole}}
&= \frac{3e^2}{2\pi}\sum_{f}\left(\dfrac{q_{\sss{f}}}{e}\right)^2 (1-e^{\beta p_0})\int_{0}^{\infty} dk\,k^2 \int_{-1}^{1}d\xi\Bigg[\zlp{2+}\nf(\olp)\nf(k)\delta(p_0-\olp-k) \nn
& +\zlm{2+}\nf(\olm)\nf(k)\delta(p_0-\olm-k)+\zrp{2+}\nf(\orp)\nf(k)\delta(p_0-\orp-k)  \nn
&+\zrm{2+}\nf(\orm)\nf(k)\delta(p_0-\orm-k)-\zlp{2+}\nf(\olp)\nf(-k)\delta(p_0-\olp+k) \nn
&-\zlm{2+}\nf(\olm)\nf(-k)\delta(p_0-\olm+k)-\zrp{2+}\nf(\orp)\nf(-k)\delta(p_0-\orp+k) \nn
&-\zlp{2+}\nf(\olp)\nf(-k)\delta(p_0-\olp+k)\Bigg].
\end{align}
Also using  Eq.~\eqref{rho^0_0} in Eq.~\eqref{spec_pimumu3}, we obtain
\begin{align}
\left.\mathrm{Im}\indices{^{3\mu}_{\mu}} \right|_{\text{pole-pole}}&= \frac{3e^2}{2\pi}\sum_{f}\left(\dfrac{q_{\sss{f}}}{e}\right)^2 
(1-e^{\beta p_0})\int_{0}^{\infty} dk\,k^2 \int_{-1}^{1}d\xi\,\xi\Bigg[\zlp{3+}\nf(\olp)\nf(k)\delta(p_0-\olp-k)  \nn
&+\zlm{3+}\nf(\olm)\nf(k)\delta(p_0-\olm-k)-\zrp{3+}\nf(\orp)\nf(k)\delta(p_0-\orp-k)  \nn
&-\zrm{3+}\nf(\orm)\nf(k)\delta(p_0-\orm-k)-\zlp{3+}\nf(\olp)\nf(-k)\delta(p_0-\olp+k)  \nn
&-\zlm{3+}\nf(\olm)\nf(-k)\delta(p_0-\olm+k)-\zrp{3+}\nf(\orp)\nf(-k)\delta(p_0-\orp+k) \nn
&-\zrm{3+}\nf(\orm)\nf(-k)\delta(p_0-\orm+k)\Bigg].
\end{align}
\noindent{\sf{(b) Magnetic field dependent part:}}\\

We begin by stating that some  terms with derivatives of Dirac $\delta$ functions are present. But after doing integration by parts, these terms will eventually get eliminated. Also, using the parity properties of the $\delta$ function and its derivatives it is easy to see that $\rho^{1}_{(0)}(-\om^{\p})=-\rho^{1}_{(0)}(\om^{\p})$.
Using Eq.~\eqref{rho^0_1} in Eq.~\eqref{spec_pimumu4}, we get
\begin{align}
\left.\mathrm{Im}\Pi\indices{^{4\mu}_{\mu}}\right|_{\text{pole-pole}} &=-\frac{3e^2}{4\pi}\sum_{f}\left(\dfrac{q_{\sss{f}}}{e}\right)^2 (1-e^{\beta p_0})q_{\sss{f}}B\int_{0}^{\infty}dk\,k^3\int_{-1}^{1}d\xi\,\xi\int_{-\infty}^{\infty}d\om^{\p} \nf(\om^{\p})\nf(p_0-\om^{\p}) \nn
&\hspace{4cm}\times[\rho_{L}^{1}(p_0-\om^{\p})-\rho_{R}^{1}(p_0-\om^{\p})]\rho_{1}^{(0)}(\om^{\p}) \nn
&=-\frac{3e^2}{4\pi}\sum_{f}\left(\dfrac{q_{\sss{f}}}{e}\right)^2 (1-e^{\beta p_0})q_{\sss{f}}B\int_{0}^{\infty}dk\,\int_{-1}^{1}d\xi\,\xi\int_{-\infty}^{\infty}d\om^{\p} \nf(\om^{\p})\nf(p_0-\om^{\p}) \nn
&\times\left[\rho_{L}^{1}(p_0-\om^{\p})-\rho_{R}^{1}(p_0-\om^{\p})\right]\left[\delta(\om^{\p}-k)-\delta(\om^{\p}+k)+k\dfrac{\partial}{\partial \om^{\p}}\left(\delta(\om^{\p}-k)+\delta(\om^{\p}+k)\right)\right] \nn
&=-\frac{3e^2}{4\pi}\sum_{f}\left(\dfrac{q_{\sss{f}}}{e}\right)^2 (1-e^{\beta p_0})q_{\sss{f}}B\int_{0}^{\infty}dk\,\int\limits_{-1}^{1}\!d\xi\,\xi\Bigg[\nf(k)\nf(p_0-k)\Big\lbrace\rho_{L}^{1}(p_0-k)-\rho_{R}^{1}(p_0-k)\Big\rbrace  \nn
&-\nf(-k)\nf(p_0+k)\Big\lbrace\rho_{L}^{1}(p_0+k)-\rho_{R}^{1}(p_0+k)\Big\rbrace +k\int_{-\infty}^{\infty}d\om\,\nf(\om)\nf(p_0-\om)  \nn
&\hspace{4cm}\times\Bigg(\rho_{L}^{1}(p_0-\om)-\rho_{R}^{1}(p_0-\om)\Bigg)\Bigg(\delta^{\p}(\om-k)+\delta^{\p}(\om+k)\Bigg)\Bigg].
\end{align} 
At this point, we use partial fraction method to eliminate $\delta^{\p}(\om\pm k)$, and it gives
\begin{align}
\left.\mathrm{Im} \Pi\indices{^{4\mu}_{\mu}}\right|_{\text{pole-pole}} &= -\frac{3e^2}{4\pi}\sum_{f}\left(\dfrac{q_{\sss{f}}}{e}\right)^2 (1-e^{\beta p_0})q_{\sss{f}}B\int_{0}^{\infty}dk\,\int_{-1}^{1}d\xi\,\xi\Bigg[\nf(k)\nf(p_0-k)\Big\lbrace\rho_{L}^{1}(p_0-k)-\rho_{R}^{1}(p_0-k)\Big\rbrace \nn
&-\nf(-k)\nf(p_0+k)\Big\lbrace\rho_{L}^{1}(p_0+k)-\rho_{R}^{1}(p_0+k)\Big\rbrace \nn
&-k\int_{-\infty}^{\infty}d\om\,\dfrac{\partial }{\partial \om}\Bigg\lbrace\nf(\om)\nf(p_0-\om)[\rho_{L}^{1}(p_0-\om)-\rho_{R}^{1}(p_0-\om)]\Bigg\rbrace\Bigg(\delta(\om-k)+\delta(\om+k)\Bigg)\Bigg] \nn
&=-\frac{3e^2}{4\pi}\sum_{f}\left(\dfrac{q_{\sss{f}}}{e}\right)^2 (1-e^{\beta p_0})q_{\sss{f}}B\int_{0}^{\infty}dk\,\int_{-1}^{1}d\xi\,\xi\Bigg[\nf(k)\nf(p_0-k)\Big\lbrace\rho_{L}^{1}(p_0-k)-\rho_{R}^{1}(p_0-k)\Big\rbrace \nn
&-\nf(-k)\nf(p_0+k)\Big\lbrace\rho_{L}^{1}(p_0+k)-\rho_{R}^{1}(p_0+k)\Big\rbrace-k\dfrac{\partial}{\partial k}\Bigg(\nf(k)\nf(p_0-k)\Big\lbrace\rho_{L}^{1}(p_0-k) \nn
&-\rho_{R}^{1}(p_0-k)\Big\rbrace-\nf(-k)\nf(p_0+k)\Big\lbrace\rho_{L}^{1}(p_0+k)-\rho_{R}^{1}(p_0+k)\Big\rbrace\Bigg)\Bigg] \nn
&=-\frac{3e^2}{4\pi}\sum_{f}\left(\dfrac{q_{\sss{f}}}{e}\right)^2 (1-e^{\beta p_0})q_{\sss{f}}B\int_{0}^{\infty}dk\,\int_{-1}^{1}d\xi\,\xi\Bigg[2\nf(k)\nf(p_0-k)\Big\lbrace\rho_{L}^{1}(p_0-k)-\rho_{R}^{1}(p_0-k)\Big\rbrace \nn
&-\nf(-k)\nf(p_0+k)\Big\lbrace\rho_{L}^{1}(p_0+k)-\rho_{R}^{1}(p_0+k)\Big\rbrace-\dfrac{\partial}{\partial k}\Bigg(k\nf(k)\nf(p_0-k)\Big\lbrace\rho_{L}^{1}(p_0-k) \nn
&-\rho_{R}^{1}(p_0-k)\Big\rbrace-k\nf(-k)\nf(p_0+k)\Big\lbrace\rho_{L}^{1}(p_0+k)-\rho_{R}^{1}(p_0+k)\Big\rbrace\Bigg)\Bigg].
\end{align}
The last term, i.e., the term that contains a derivative with respect to $k$, when integrated out gives the  boundary term and it vanishes. Also, by using the properties of the $\delta$ function, one obtains the pole-pole part as
\begin{align}
\left.\mathrm{Im} \Pi\indices{^{4\mu}_{\mu}}\right|_{\text{pole-pole}} &=-\frac{3e^2}{2\pi}\sum_{f}\left(\dfrac{q_{\sss{f}}}{e}\right)^2 
(1-e^{\beta p_0})q_{\sss{f}}B\int_{-1}^{1}d\xi\,\xi\int_{0}^{\infty}dk\Bigg[\nf(k)\Bigg\lbrace\zlp{1+}\nf(\olp)\delta(p_0-k-\olp)  \nn
&+\zlm{1+}\nf(\olm)\delta(p_0-k-\olm)-\zrp{1+}\nf(\orp)\delta(p_0-k-\orp)-\zrm{1+}\nf(\orm)\delta(p_0-k-\orm) \Bigg\rbrace \nn
&-\nf(-k)\Bigg\lbrace\zlp{1+}\nf(\olp)\delta(p_0+k-\olp)+\zlm{1+}\nf(\olm)\delta(p_0+k-\olm) \nn
&-\zrp{1+}\nf(\orp)\delta(p_0+k-\orp)-\zrm{1+}\nf(\orm)\delta(p_0+k-\orm) \Bigg\rbrace \Bigg].
\end{align}
Using \eqref{rho^1_1} in Eq.~\eqref{spec_pimumu5}, we get
\begin{align}
\left.\mathrm{Im} \Pi\indices{^{5\mu}_{\mu}}\right|_{\text{pole-pole}} &=\frac{3e^2}{4\pi}\sum_{f}\left(\dfrac{q_{\sss{f}}}{e}\right)^2 
(1-e^{\beta p_0})q_{\sss{f}}B\int_{-1}^{1}d\xi\,\xi\int_{0}^{\infty}dk\Bigg[\nf(k)\Bigg\lbrace\zlp{2+}\nf(\olp)\delta(p_0-k-\olp)  \nn
&+\zlm{2+}\nf(\olm)\delta(p_0-k-\olm)-\zrp{2+}\nf(\orp)\delta(p_0-k-\orp)-\zrm{2+}\nf(\orm)\delta(p_0-k-\orm) \Bigg\rbrace \nn
&+\nf(-k)\Bigg\lbrace\zlp{2+}\nf(\olp)\delta(p_0+k-\olp)+\zlm{2+}\nf(\olm)\delta(p_0+k-\olm) \nn
&-\zrp{2+}\nf(\orp)\delta(p_0+k-\orp)-\zrm{2+}\nf(\orm)\delta(p_0+k-\orm) \Bigg\rbrace \Bigg].
\end{align}
Finally using Eq.~\eqref{rho^1_1} in Eq.~\eqref{spec_pimumu6}, we get
\begin{align}
\left.\mathrm{Im} \Pi\indices{^{6\mu}_{\mu}}\right|_{\text{pole-pole}} &=\frac{3e^2}{4\pi}\sum_{f}\left(\dfrac{q_{\sss{f}}}{e}\right)^2 
(1-e^{\beta p_0})q_{\sss{f}}B\int_{-1}^{1}d\xi\int_{0}^{\infty}dk\Bigg[\nf(k)\Bigg\lbrace\zlp{2+}\nf(\olp)\delta(p_0-k-\olp)  \nn
&+\zlm{2+}\nf(\olm)\delta(p_0-k-\olm)+\zrp{2+}\nf(\orp)\delta(p_0-k-\orp)+\zrm{2+}\nf(\orm)\delta(p_0-k-\orm) \Bigg\rbrace \nn
&+\nf(-k)\Bigg\lbrace\zlp{2+}\nf(\olp)\delta(p_0+k-\olp)+\zlm{2+}\nf(\olm)\delta(p_0+k-\olm) \nn
&+\zrp{2+}\nf(\orp)\delta(p_0+k-\orp)+\zrm{2+}\nf(\orm)\delta(p_0+k-\orm) \Bigg\rbrace \Bigg].
\end{align}

\noindent{\sf (c) Dilepton rate from various processes in pole-pole part in presence of magnetic field:}\\

\begin{figure}[tbh]
\subfigure{
	\includegraphics[scale=0.69]{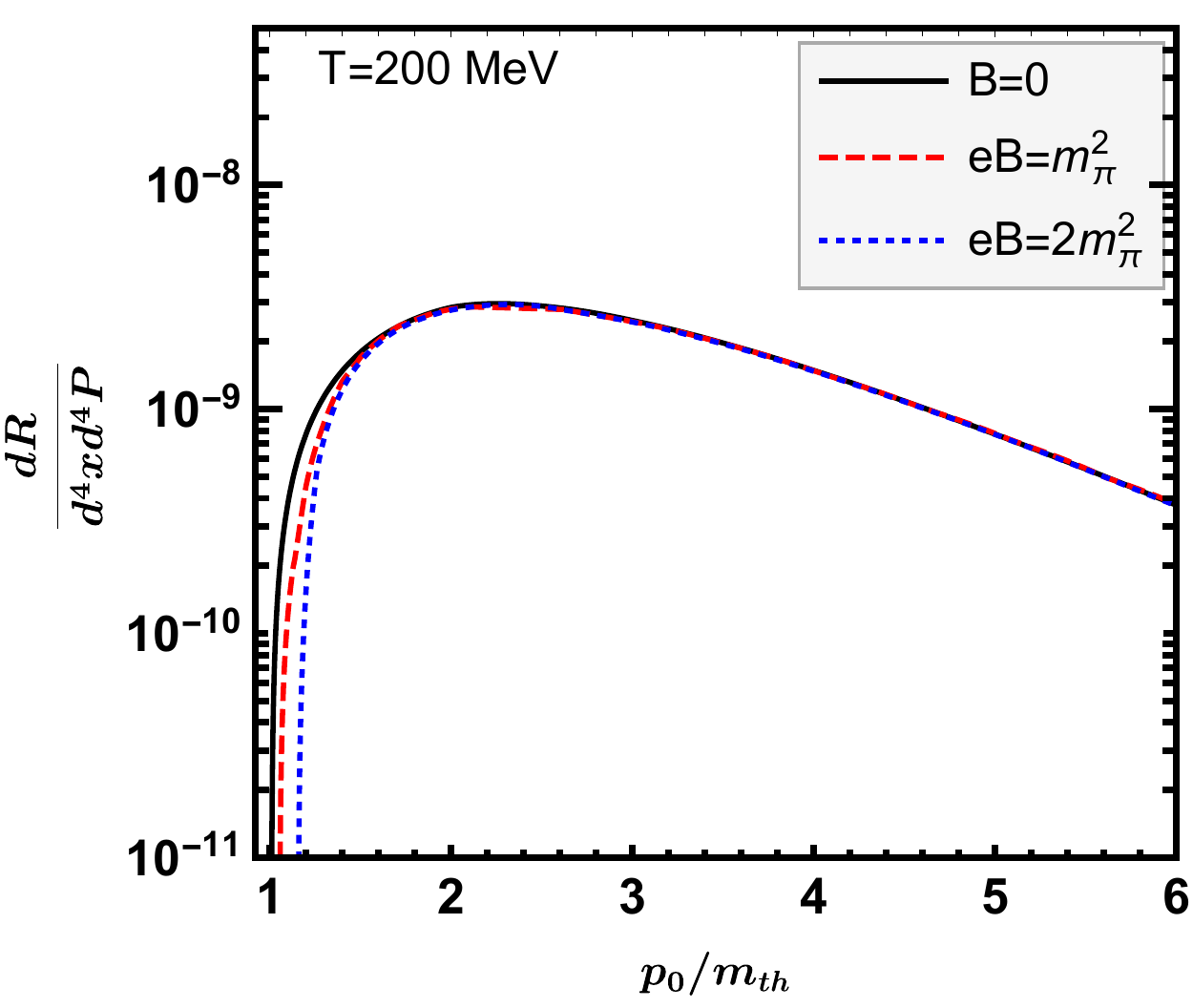}}
\subfigure{
\includegraphics[scale=0.69]{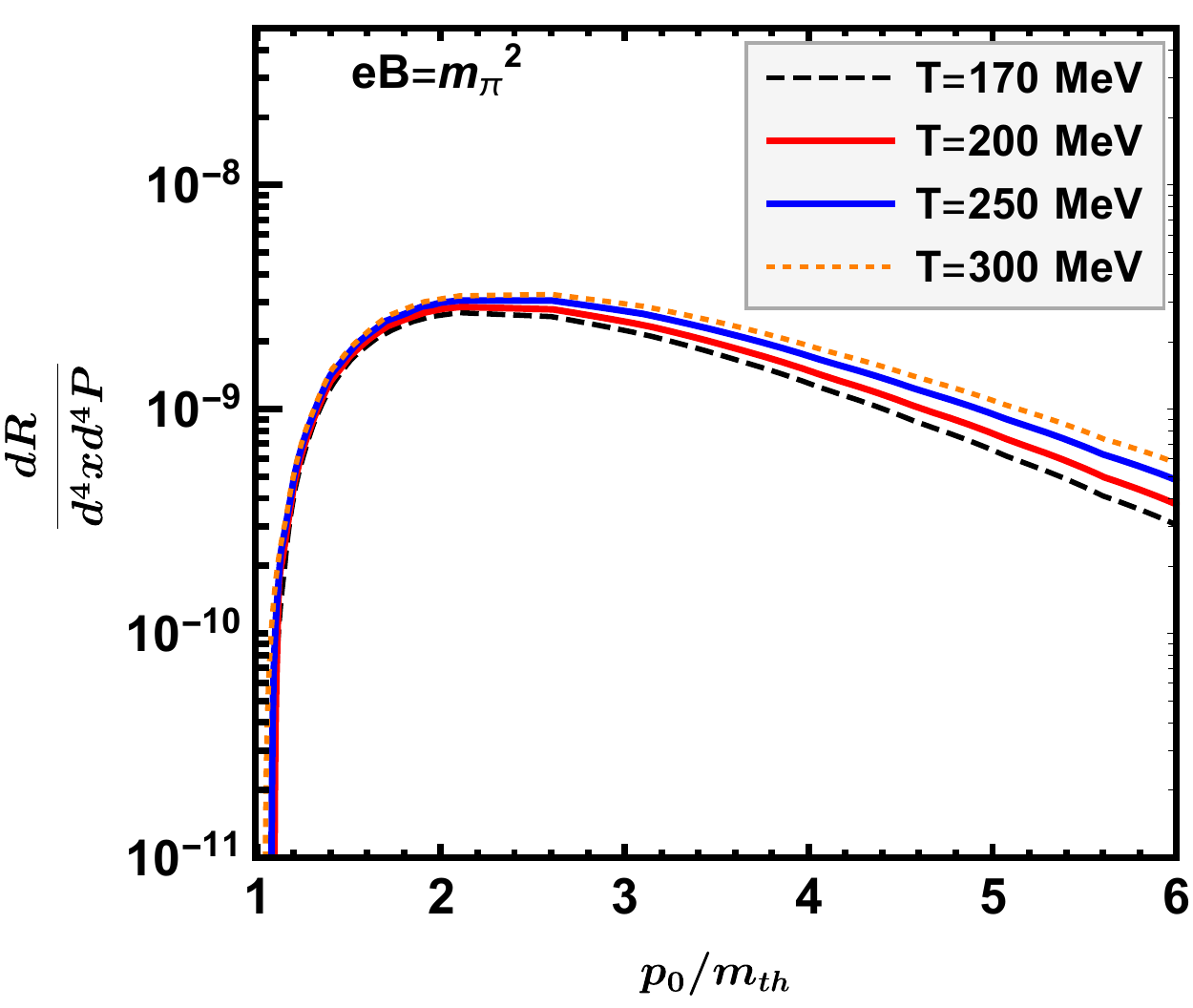}}
\caption{Pole-pole contribution of the dilepton production rate as a function of the energy of dilepton in the center-of-mass reference frame at $T=200$ MeV with different magnetic field (left panel) and $eB=m_\pi^2$  with different temperature (right panel). }
	\label{fig:pp}
\end{figure}

We note that for numerical computation we change the integration from spherical polar to cylindrical polar 
through the transformation $k_{\sss{\perp}}=k \sqrt{1-\xi^2},\,\,k_z=k \xi$, where $\xi=\cos\theta$. Using \eqref{pimumu_together} and grouping the delta 
functions together we get the dilepton rates in terms of the cylindrical polar coordinate from various processes discussed in Sec.~\ref{sec:disp}  as follows:
\begin{enumerate}
\item $\underline{q_{L(+)}q\longrightarrow \gamma^{*}}$
\begin{align}
\left.\frac{dR}{d^4xd^4P}\right|^{q_{L(+)}q\rightarrow \gamma^*}&= \frac{\alpha^2}{2p^2_0\pi^4}\sum_{f}\left(\dfrac{q_{\sss{f}}}{e}\right)^2\int_{0}^{\infty}dk_{\perp}\,\,k_{\perp}\int_{-\infty}^{\infty}dk_z n_F\Big(p_0-\sqrt{k^2_{\perp}+k^2_z}\Big)n_F\Big(\sqrt{k^2_{\perp}+k^2_z}\Big)\nn
&\times\Bigg[Z^1_{L(+)}+Z^2_{L(+)}  +\frac{k_z}{\sqrt{k^2_{\perp}+k^2_{z}}}Z^3_{L(+)}+\frac{q_fB}{k^2_{\perp}+k^2_{z}}\Bigg(\frac{k_z}{\sqrt{k^2_{\perp}+k^2_{z}}}Z^1_{L(+)}\nn
&+\frac{k_z}{2\sqrt{k^2_{\perp}+k^2_{z}}}Z^2_{L(+)}+\frac{1}{2}Z^3_{L(+)}\Bigg)\Bigg]
\delta\left(p_0-\omega_{L(+)}(k_{\perp},k_z)-\sqrt{k^2_{\perp}+k^2_{z}}\right).
\end{align}

\item {$\underline{q_{L(-)}q\longrightarrow \gamma^{*}}$}
\begin{align}
\left.\frac{dR}{d^4xd^4P}\right|^{q_{L(-)}q\rightarrow \gamma^*}&= \frac{\alpha^2}{2p^2_0\pi^4}\sum_{f}\left(\dfrac{q_{\sss{f}}}{e}\right)^2\int_{0}^{\infty}dk_{\perp}\,\,k_{\perp}\int_{-\infty}^{\infty}dk_z n_F\Big(p_0-\sqrt{k^2_{\perp}+k^2_z}\Big)n_F\Big(\sqrt{k^2_{\perp}+k^2_z}\Big)\nn
&\times\Bigg[Z^1_{L(-)}+Z^2_{L(-)}   
+\frac{k_z}{\sqrt{k^2_{\perp}+k^2_{z}}}Z^3_{L(-)}+\frac{q_fB}{k^2_{\perp}+k^2_{z}}\Bigg(\frac{k_z}{\sqrt{k^2_{\perp}+k^2_{z}}}Z^1_{L(-)}\nn
&+\frac{k_z}{2\sqrt{k^2_{\perp}+k^2_{z}}}Z^2_{L(-)}+\frac{1}{2}Z^3_{L(-)}\Bigg)\Bigg]\delta\Big(p_0-\omega_{L(-)}(k_{\perp},k_z)-\sqrt{k^2_{\perp}+k^2_{z}}\Big).
\end{align}
\item {$\underline{q_{R(+)}q\longrightarrow \gamma^{*}}$}
\begin{align}
\left.\frac{dR}{d^4xd^4P}\right|^{q_{R(+)}q\rightarrow \gamma^*} &= \frac{\alpha^2}{2p^2_0\pi^4}\sum_{f}\left(\dfrac{q_{\sss{f}}}{e}\right)^2\int_{0}^{\infty}dk_{\perp}\,\,k_{\perp}\int_{-\infty}^{\infty}dk_z n_F\Big(p_0-\sqrt{k^2_{\perp}+k^2_z}\Big)n_F\Big(\sqrt{k^2_{\perp}+k^2_z}\Big)\nn
&\times\Bigg[Z^1_{R(+)}+Z^2_{R(+)}    -\frac{k_z}{\sqrt{k^2_{\perp}+k^2_{z}}}Z^3_{R(+)}-\frac{q_fB}{k^2_{\perp}+k^2_{z}}\Bigg(\frac{k_z}{\sqrt{k^2_{\perp}+k^2_{z}}}Z^1_{R(+)}\nn
&+\frac{k_z}{2\sqrt{k^2_{\perp}+k^2_{z}}}Z^2_{R(+)}-\frac{1}{2}Z^3_{R(+)}\Bigg)\Bigg]\delta\Big(p_0-\omega_{R(+)}(k_{\perp},k_z)-\sqrt{k^2_{\perp}+k^2_{z}}\Big).
\end{align}

\item {$\underline{q_{R(-)}q\longrightarrow \gamma^{*}}$}
\begin{align}
\left.\frac{dR}{d^4xd^4P}\right|^{q_{R(-)}q\rightarrow \gamma^*}&= \frac{\alpha^2}{2p^2_0\pi^4}\sum_{f}\left(\dfrac{q_{\sss{f}}}{e}\right)^2\int_{0}^{\infty}dk_{\perp}\,\,k_{\perp}\int_{-\infty}^{\infty}dk_z n_F\Big(p_0-\sqrt{k^2_{\perp}+k^2_z}\Big)n_F\Big(\sqrt{k^2_{\perp}+k^2_z}\Big)\nn
&\times\Bigg[Z^1_{R(-)}+Z^2_{R(-)}-\frac{k_z}{\sqrt{k^2_{\perp}+k^2_{z}}}Z^3_{R(-)}-\frac{q_fB}{k^2_{\perp}+k^2_{z}}\Bigg(\frac{k_z}{\sqrt{k^2_{\perp}+k^2_{z}}}Z^1_{R(-)}\nn
&+\frac{k_z}{2\sqrt{k^2_{\perp}+k^2_{z}}}Z^2_{R(-)}-\frac{1}{2}Z^3_{R(-)}\Bigg)\Bigg]\delta\Big(p_0-\omega_{R(-)}(k_{\perp},k_z)-\sqrt{k^2_{\perp}+k^2_{z}}\Big).
\end{align}
\item {$\underline{q_{L(+)}\longrightarrow q\gamma^{*}}$}
\begin{align}
\left.\frac{dR}{d^4xd^4P}\right|^{q_{L(+)}\rightarrow q\gamma^*}&= \frac{\alpha^2}{2p^2_0\pi^4}\sum_{f}\left(\dfrac{q_{\sss{f}}}{e}\right)^2\int_{0}^{\infty}dk_{\perp}\,\,k_{\perp}\int_{-\infty}^{\infty}dk_z n_F\Big(p_0+\sqrt{k^2_{\perp}+k^2_z}\Big)n_F\Big(-\sqrt{k^2_{\perp}+k^2_z}\Big)\nonumber\\
&\times\Bigg[Z^1_{L(+)}-Z^2_{L(+)}-\frac{k_z}{\sqrt{k^2_{\perp}+k^2_{z}}}Z^3_{L(+)}-\frac{q_fB}{k^2_{\perp}+k^2_{z}}\Bigg(\frac{k_z}{\sqrt{k^2_{\perp}+k^2_{z}}}Z^1_{L(+)}\nn
&-\frac{k_z}{2\sqrt{k^2_{\perp}+k^2_{z}}}Z^2_{L(+)}-\frac{1}{2}Z^3_{L(+)}\Bigg)\Bigg]\delta\Big(p_0-\omega_{L(+)}(k_{\perp},k_z)+\sqrt{k^2_{\perp}+k^2_{z}}\Big).
\end{align}
\item {$\underline{q_{L(-)}\longrightarrow q\gamma^{*}}$}
\begin{align}
\left.\frac{dR}{d^4xd^4P}\right|^{q_{L(-)}\rightarrow q\gamma^*} &= \frac{\alpha^2}{p^2_0\pi^4}\sum_{f}\left(\dfrac{q_{\sss{f}}}{e}\right)^2\int_{0}^{\infty}dk_{\perp}\,\,k_{\perp}\int_{-\infty}^{\infty}dk_z n_F\Big(p_0+\sqrt{k^2_{\perp}+k^2_z}\Big)n_F\Big(-\sqrt{k^2_{\perp}+k^2_z}\Big)\nn
&\times\Bigg[Z^1_{L(-)}-Z^2_{L(-)} 
-\frac{k_z}{\sqrt{k^2_{\perp}+k^2_{z}}}Z^3_{L(-)}-\frac{q_fB}{k^2_{\perp}+k^2_{z}}\Bigg(\frac{k_z}{\sqrt{k^2_{\perp}+k^2_{z}}}Z^1_{L(-)}\nn
&-\frac{k_z}{2\sqrt{k^2_{\perp}+k^2_{z}}}Z^2_{L(-)}-\frac{1}{2}Z^3_{L(-)}\Bigg)\Bigg]\delta\Big(p_0-\omega_{L(-)}(k_{\perp},k_z)+\sqrt{k^2_{\perp}+k^2_{z}}\Big).
\end{align}
\item {$\underline{q_{R(+)}\longrightarrow q\gamma^{*}}$}
\begin{align}
\left.\frac{dR}{d^4xd^4P}\right|^{q_{R(+)}\rightarrow q\gamma^*}&= \frac{\alpha^2}{2p^2_0\pi^4}\sum_{f}\left(\dfrac{q_{\sss{f}}}{e}\right)^2\int_{0}^{\infty}dk_{\perp}\,\,k_{\perp}\int_{-\infty}^{\infty}dk_z n_F\Big(p_0+\sqrt{k^2_{\perp}+k^2_z}\Big)n_F\Big(-\sqrt{k^2_{\perp}+k^2_z}\Big)\nn
&\times\Bigg[Z^1_{R(+)}-Z^2_{R(+)}
+\frac{k_z}{\sqrt{k^2_{\perp}+k^2_{z}}}Z^3_{R(+)}+\frac{q_fB}{k^2_{\perp}+k^2_{z}}\Bigg(\frac{k_z}{\sqrt{k^2_{\perp}+k^2_{z}}}Z^1_{R(+)}\nn
&-\frac{k_z}{2\sqrt{k^2_{\perp}+k^2_{z}}}Z^2_{R(+)}+\frac{1}{2}Z^3_{R(+)}\Bigg)\Bigg]\delta\Big(p_0-\omega_{R(+)}(k_{\perp},k_z)+\sqrt{k^2_{\perp}+k^2_{z}}\Big).
\end{align}
\item {$\underline{q_{R(-)}\longrightarrow q\gamma^{*}}$}
\begin{align}
\left.\frac{dR}{d^4xd^4P}\right|^{q_{R(-)}\rightarrow q\gamma^*} &= \frac{\alpha^2}{2p^2_0\pi^4}\sum_{f}\left(\dfrac{q_{\sss{f}}}{e}\right)^2\int_{0}^{\infty}dk_{\perp}\,\,k_{\perp}\int_{-\infty}^{\infty}dk_z n_F\Big(p_0+\sqrt{k^2_{\perp}+k^2_z}\Big)n_F\Big(-\sqrt{k^2_{\perp}+k^2_z}\Big)\nn
&\times\Bigg[Z^1_{R(-)}-Z^2_{R(-)}+\frac{k_z}{\sqrt{k^2_{\perp}+k^2_{z}}}Z^3_{R(-)}+\frac{q_fB}{k^2_{\perp}+k^2_{z}}\Bigg(\frac{k_z}{\sqrt{k^2_{\perp}+k^2_{z}}}Z^1_{R(-)}\nn
&-\frac{k_z}{2\sqrt{k^2_{\perp}+k^2_{z}}}Z^2_{R(-)}+\frac{1}{2}Z^3_{R(-)}\Bigg)\Bigg]\delta\Big(p_0-\omega_{R(-)}(k_{\perp},k_z)+\sqrt{k^2_{\perp}+k^2_{z}}\Big).
\end{align}
\end{enumerate}
From the parity symmetry of the dispersion mode, it is possible to show that
\bea
\left.\frac{dR}{d^4xd^4P}\right|^{\omega_{L(+)}k\rightarrow \gamma^*}&=&\left.\frac{dR}{d^4xd^4P}\right|^{\omega_{R(+)}k\rightarrow \gamma^*},\nn
\left.\frac{dR}{d^4xd^4P}\right|^{\omega_{L(-)}k\rightarrow \gamma^*}&=&\left.\frac{dR}{d^4xd^4P}\right|^{\omega_{R(-)}k\rightarrow \gamma^*},\nn
\left.\frac{dR}{d^4xd^4P}\right|^{\omega_{L(+)}\rightarrow k\gamma^*}&=&\left.\frac{dR}{d^4xd^4P}\right|^{\omega_{R(+)}\rightarrow k\gamma^*},\nn
\left.\frac{dR}{d^4xd^4P}\right|^{\omega_{L(-)}\rightarrow k\gamma^*}&=&\left.\frac{dR}{d^4xd^4P}\right|^{\omega_{R(-)}\rightarrow k\gamma^*}.
\eea
Finally, the pole-pole contribution of the hard dilepton rate becomes
\bea
\left.\frac{dR}{d^4xd^4P}\right|^{\rm{pp}}&=&2\left(\left.\frac{dR}{d^4xd^4P}\right|^{\omega_{L(+)}k\rightarrow \gamma^*} +\left.\frac{dR}{d^4xd^4P}\right|^{\omega_{L(-)}k\rightarrow \gamma^*}+\left.\frac{dR}{d^4xd^4P}\right|^{\omega_{L(+)}\rightarrow k\gamma^*} +\left.\frac{dR}{d^4xd^4P}\right|^{\omega_{L(-)}\rightarrow k\gamma^*}\right).
\label{pp_total}
\eea
We note that the various soft decay modes will contribute only to the soft dilepton production at low energy. Since we are interested in hard dilepton production rate, only the annihilation modes will contribute and we will omit those soft decay modes from our considerations. The resulting pole-pole contribution is plotted in Fig.~\ref{fig:pp}. In the left panel the rate is displayed as a function of dilepton energy at  $T=200$ MeV but for different magnetic fields. In the absence of magnetic field ($eB=0$) the annihilation between a hard and a soft quark starts when dilepton energy $E=m_{th}$ and resembles that of  $qq_+\longrightarrow \gamma^*\longrightarrow l^+l^-$ as given in Fig.~~\ref{fig:dilepton_rate_eb0}. As the magnetic field is turned on, all four quasiparticle modes, namely, $\olp$, $\olm$, $\orp$, $\orm$, as shown in Fig.~\ref{fig:HLLfig}, separately participate in annihilation with hard quark. As can be seen, the dilepton rate at finite magnetic field begins at little higher energy of the virtual photon compare to the vanishing magnetic field. This is because the presence of magnetic field contributes to the thermomagnetic mass which is lower than  the thermal mass. As the energy of the dilepton increases, the rate becomes almost equal to that  in absence of magnetic field. In the right panel of Fig.~\ref{fig:pp}, the rate is displayed for various temperatures for a given magnetic filed. At energy up to the  $E=p_0\approx 2 m_{th}$, the rate is found to be almost independent of $T$ as magnetic field may be the dominant scale there. At energies $E=p_0 > 2m_{th}$, the rate increases with the  increase of $T$ as $T$ is the dominant scale in the weak field approximation.
\subsubsection{Pole-cut contribution}
The presence 
of $\Theta$  due to spacelike momentum in the Landau cut contribution of the spectral function, $\Theta(k^2-\omega^2)\beta_{\sss{L(R)}}^{i}(\omega,k_{\sss{\perp}},k_z)$, immensely simplifies the pole-cut rate.
From Eq.~\eqref{ptc}, we get
\begin{align}
\left.\mathrm{Im}\Pi^{1\mu}_{\mu}\right|_{\text{pole-cut}} &= \frac{3e^2}{2\pi}(e^{\beta p_0}-1)\sum_{f}\left(\dfrac{q_{\sss{f}}}{e}\right)^2\int_{0}^{\infty}dk\,k^2 \int_{-1}^{1}d\xi \int_{-\infty}^{\infty}d\om\Theta(k^2-\om^2)\left[\beta_{L}^{1}(\om)+\beta_{R}^{1}(\om)\right]n_{\sss{F}}(\om) \nn
&\hspace{6cm}\times\left[\nf(-k)\delta(p_0-\om+k)+\nf(k)\delta(p_0-\om-k)\right] .
\end{align} 
We note that the term with $\delta(p_0-\om+k)$ will have no contribution because 
$\Theta[k^2-(p_0+ k)^2]=\Theta[-p_0(p_0+2k)^2]$ will never be satisfied since $k,p_0 > 0.$ The expression to evaluate the pole-cut contribution is 
\begin{align}
\left.\frac{dR}{d^4xd^4p}\right|_{\text{pole-cut}} &= \frac{\alpha^2}{2\pi^4p^2_0}\sum_{f}\left(\dfrac{q_{\sss{f}}}{e}\right)^2
\int_{-1}^{1}d\xi\int_{0}^{\infty}dk\,\,n_F(k)n_F(p_0-k)\Theta\(2k-p_0\) \nn
 &\hspace{-2cm}\times\Big[k^2\Big(\beta_L^1+\beta_R^1+\beta_L^2+\beta_R^2+\xi(\beta_L^3-\beta_R^3)\Big)
 +q_fB\Big(\xi(\beta_L^1-\beta_R^1)+\frac{1}{2}\xi\(\beta_L^2-\beta_R^2\)+\frac{1}{2}\(\beta_L^3+\beta_R^3\)\Big)\Big],
 \label{cp_total}
\end{align}
where
$\beta^{i}_{(L/R)}\equiv \beta^{i}_{(L/R)}(p_0-k,k_{\sss{\perp}},k^3)$.
 \begin{figure}[h!]
	\centering
	\includegraphics[scale=0.7]{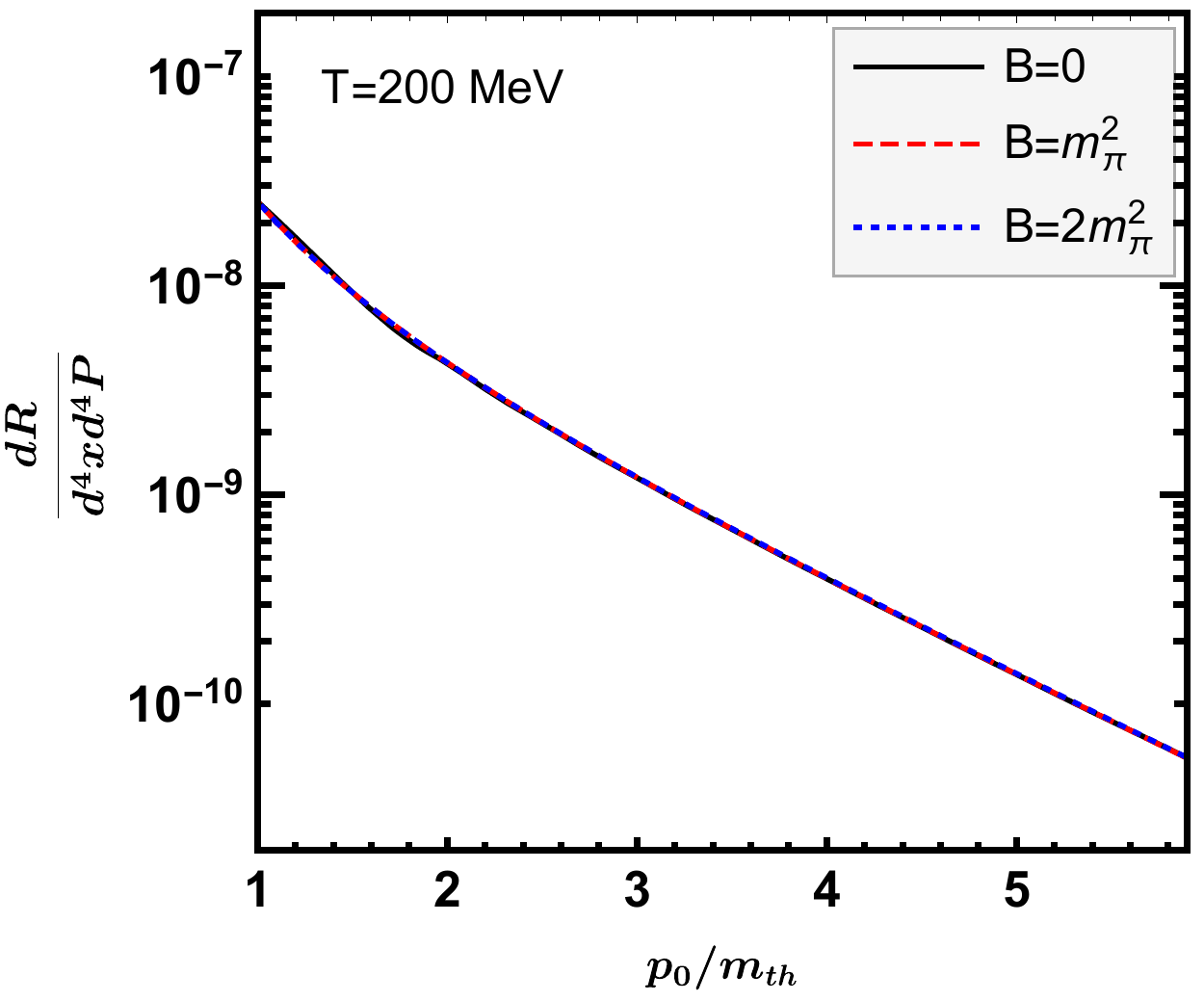}
	\includegraphics[scale=0.7]{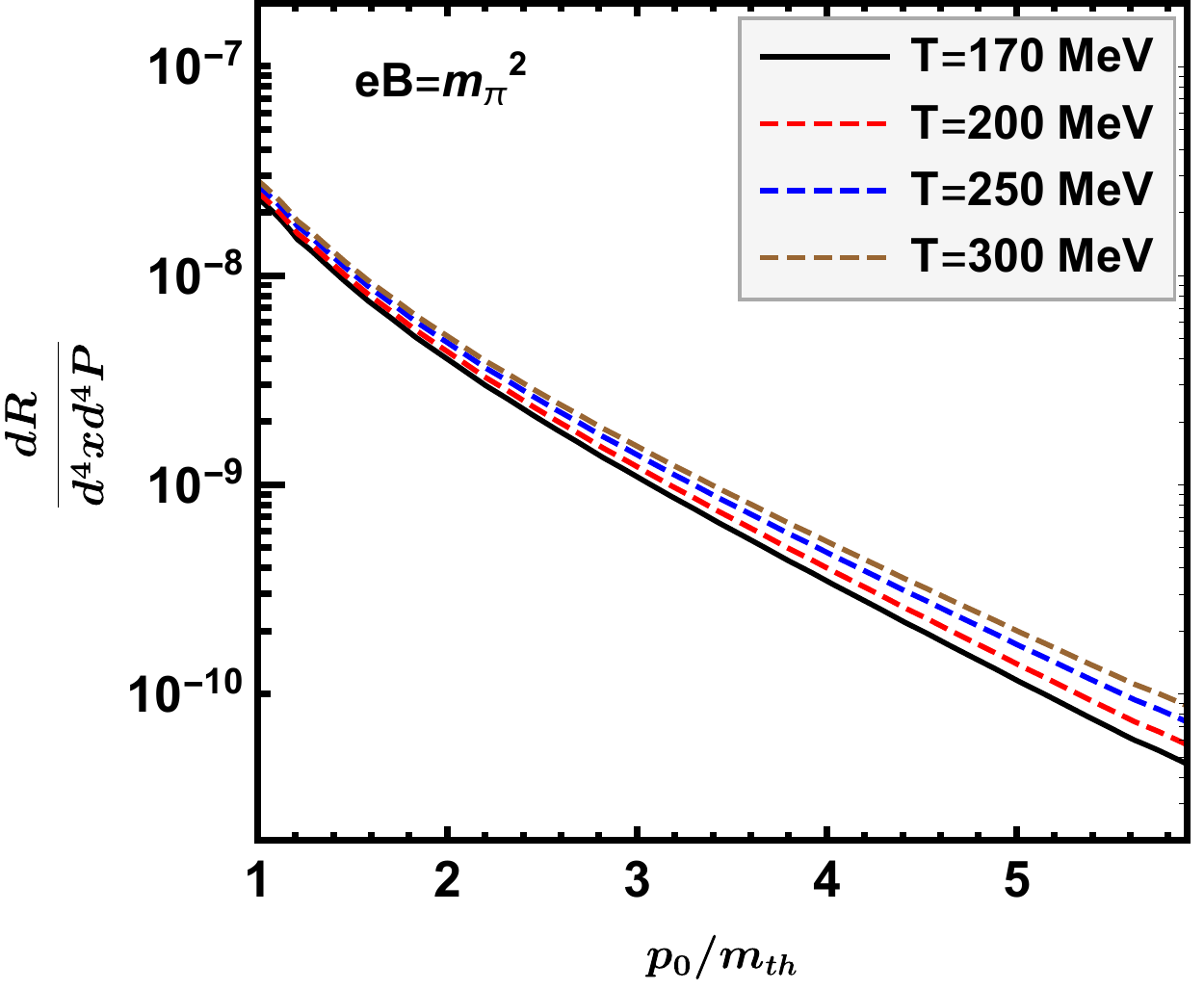}
	\caption{Same as Fig.~\ref{fig:pp} but for the pole-cut contribution.}
	\label{fig:cut_pole}
 \end{figure}
 
\begin{figure}[tbh]
	\centering
	\includegraphics[scale=0.7]{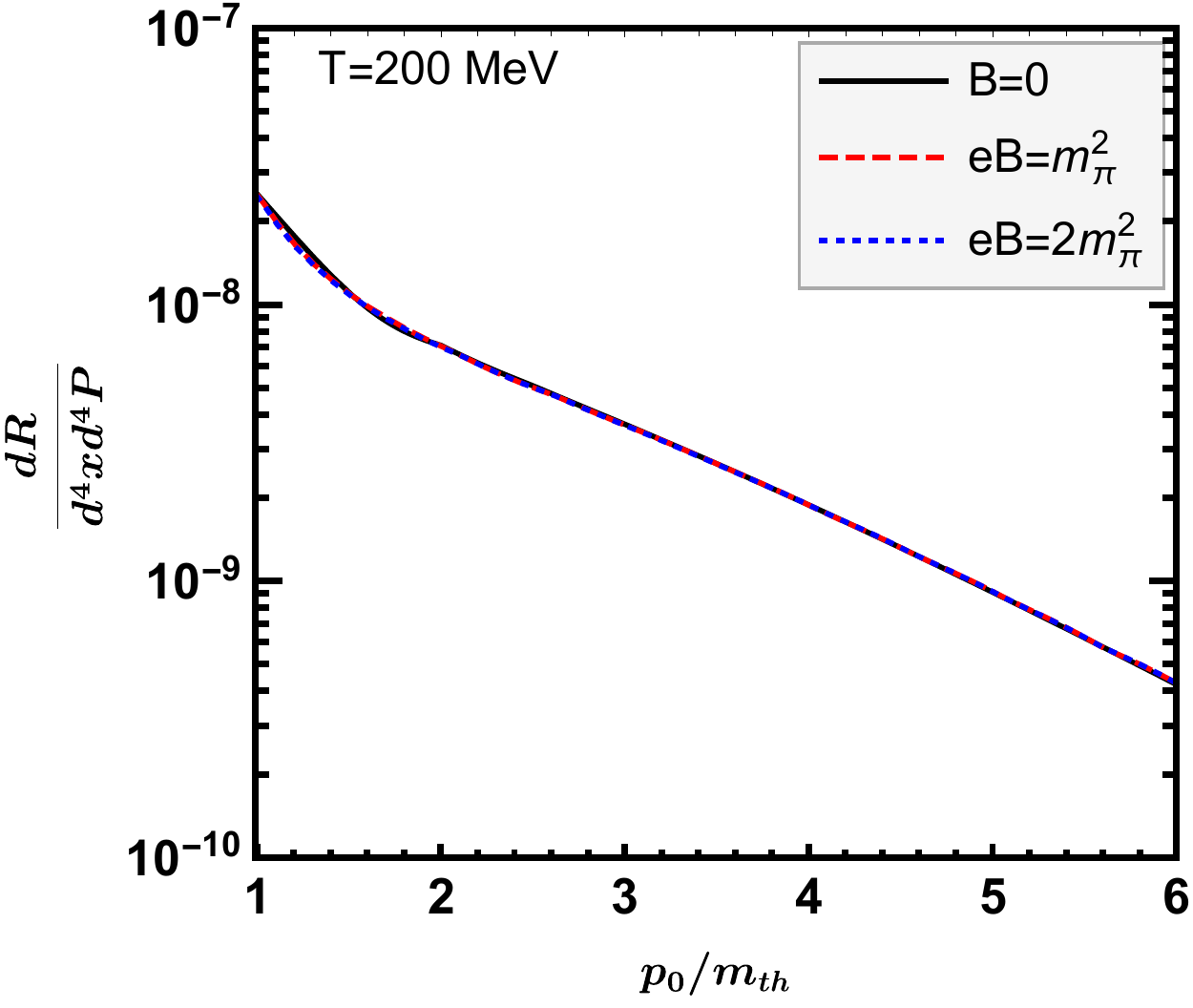}
	\includegraphics[scale=0.7]{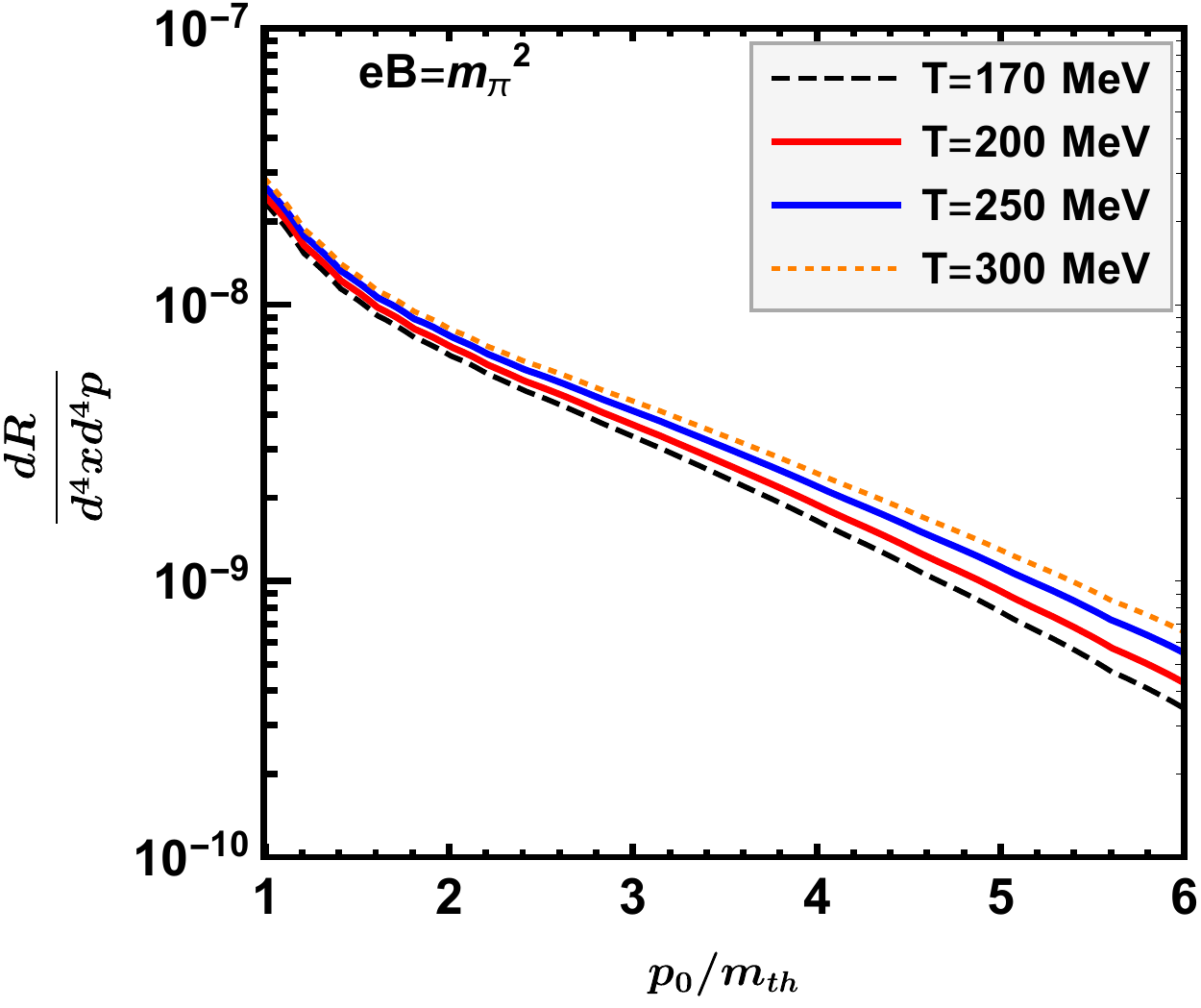}
	\caption{Total rate, sum of pole-pole and pole-cut contributions, of dilepton production r as a function of the energy of dilepton  for various magnetic fields (left panel) and for various temperatures (right panel).}
	\label{fig:total}
\end{figure}

In the left panel of Fig.~\ref{fig:cut_pole}, the pole-cut contribution is plotted for various magnetic fields with $T=200$ MeV. It is found to be independent of of the magnetic field. This is because magnetic field appears as a correction in the weak field approximation and we have considered the rate up to ${\cal O}[(eB)]$. On the other hand, in the left panel of Fig.~\ref{fig:cut_pole}, it is plotted for various temperatures for a given magnetic field. The rate is found to be enhanced with the increase in temperature as the temperature is the dominant scale in the weak field approximation. Total dilepton rate is obtained by adding the pole-pole contribution from Eq.~\eqref{pp_total} and the pole-cut contribution from Eq.~\eqref{cp_total} and is plotted in Fig.~\eqref{fig:total} with similar behavior as in Fig.~\ref{fig:pp}.
 
\section{Conclusion}
\label{concl}
In this paper, we have systematically investigated thermal dilepton production from a hot magnetized QCD medium in the weak field approximation. Since we are interested in the hard dilepton rate, it is sufficient to use just one resummed and one bare propagator in the presence of magnetic field in the photon polarization tensor diagram in Fig.~\ref{fig:dilepton_production}. We note that the earlier works were carried out using free propagators for both  the fermions in the loop in the presence of magnetic field. Since we have one resummed propagator, its spectral representation contains a pole and a (Landau) cut contribution. On the other hand, a hard spectral function corresponding to bare propagator has only pole contribution. The dilepton rate contains two types of contributions: pole-pole and pole-cut. As the magnetic field is turned on, all four quasiquark modes, namely, $\olp$, $\olm$, $\orp$, and $\orm$ individually participate in annihilation with a hard quark and contribute to the pole-pole part of the dilepton production. These annihilation processes start at higher energies as the thermomagnetic mass increases in the presence of magnetic field. The pole-cut contribution is found to dominate over those annihilation processes at low energies.

In weak field approximation magnetic field appears as a correction to the thermal contributions. Since, for simplicity, we have considered only ${\cal O}[(eB)]$ correction,  the effect of magnetic field on the rate is found to be very marginal here. For having a moderate effect of the magnetic field, one may need to take into account QCD corrections. On the other hand, one may consider a photon self-energy diagram with two resummed quark propagators along with two effective three-point vertices. In addition, a four-point vertex diagram will also contribute. This altogether will present a complete picture of soft dilepton production in one-loop order. We also note that in this calculation we have only considered the case in which the quarks are affected by the presence of the magnetic field, whereas the leptons remain unaffected as they are assumed to be produced at the edge of the fireball. Since dileptons are produced at every stage of the fireball, one should also take into account the modification of the leptons in the presence of magnetic field. All these are very interesting prospects but will indeed be very involved calculations.

\section{Acknowledgment}
The authors would like to acknowledge Aritra Bandyopadhyay and Arghya Mukherjee for useful discussions. P.~K.~R. was funded by Department of Atomic Energy (DAE), India via the project A Large Ion Collider Experiment(ALICE)/Saha Institute of Nuclear Physics(SINP). A.~D. was funded by DAE/ALICE/SINP and partially by School of Physical Sciences (SPS), National Institute of Science Education and Research (NISER). N.~H. was funded by DAE, M.~G.~M. was funded by DAE via project TPAES. 

\appendix
\section{Spectral Representation of weak field propagator upto $ \mathcal{O}(q_{f}B) $}
We need to find the spectral representation of $ S_B(K) $. To do this we write~\cite{Chyi:1999fc}
\begin{align*}
S_B(K)&=\frac{\slashed{K}}{K^2-m^2_f}+i\gamma\indices{^1}\gamma\indices{^2}\frac{\slashed{k}_{\sss{\parallel}}}{(K^2-m^2_f)^2}q_fB\\
&=\frac{\slashed{K}}{K^2-m^2_f}-\gamma_5\frac{k_0\gamma\indices{^3}-k^3\gamma\indices{_0}}{(K^2-m^2_f)^2}q_fB\\
&=\frac{k\indices{_0}}{k^2_0-\omega^2_k}\gamma\indices{^0}-|\vec{k}|\frac{1}{k^2_0-\omega^2_k}\hat{k}.\gamma-\gamma_5\left[\frac{k\indices{_0}}{(k^2_0-\omega^2_k)^2}\gamma\indices{^3}-\frac{1}{(k^2_0-\omega^2_k)^2}k^3\gamma\indices{_0}\right]q_fB.
\end{align*}
We define the spectral functions as follows
\begin{align}
\rho_0^{(1)}(k_0,|\vec{k}|)&=\frac{1}{\pi}\mbox{Im}\,f_0^{(1)}(k_0+i\epsilon,|\vec{k}|)=\frac{1}{\pi}\mbox{Im}\,\frac{k_0+i\epsilon}{(k_0+i\epsilon)^2-\omega_k^2} \label{f_0^1}.\\
\rho_0^{(0)}(k_0,|\vec{k}|)&=\frac{1}{\pi}\mbox{Im}\,f_0^{(0)}(k_0+i\epsilon,|\vec{k}|)=\frac{1}{\pi}\mbox{Im}\,\frac{1}{(k_0+i\epsilon)^2-\omega_k^2} \label{f_0^0},\\
\rho_1^{(1)}(k_0,|\vec{k}|)&=\frac{1}{\pi}\mbox{Im}\,f_1^{(1)}(k_0+i\epsilon,|\vec{k}|)=\frac{1}{\pi}\mbox{Im}\,\frac{k_0+i\epsilon}{[(k_0+i\epsilon)^2-\omega_k^2]^2} \label{f_1^1}, \\
\rho_1^{(0)}(k_0,|\vec{k}|)&=\frac{1}{\pi}\mbox{Im}\,f_1^{(0)}(k_0+i\epsilon,|\vec{k}|)=\frac{1}{\pi}\mbox{Im}\,\frac{1}{[(k_0+i\epsilon)^2-\omega_k^2]^2} \label{f_1^0}.
\end{align}
Now to prove this we need to use \cite{Das:1997gg}
\begin{align}
\lim\limits_{\epsilon\rightarrow 0}\,\,\mbox{Im}\,\frac{1}{x+i\epsilon}&=-\pi\delta(x) \label{theo1},\\
\lim\limits_{\epsilon\rightarrow 0}\,\,\mbox{Im}\,\frac{1}{(x+i\epsilon)^2}&=\pi\delta^{\prime}(x), \label{theo2}
\end{align}
where $ x,\epsilon \in \mathbb{R} $, $ \epsilon>0 $.\\ \\
To prove \eqref{theo1} and \eqref{theo2}, we use the following limiting representation of Dirac delta function
\begin{align}
\lim\limits_{\epsilon\rightarrow 0}\frac{\epsilon}{x^2+\epsilon^2}=\pi\delta(x). \label{delrep1}
\end{align}
Taking derivative with respect to $x$ on both sides of equation \eqref{delrep1}, we get 
\begin{align}
\lim\limits_{\epsilon\rightarrow 0}\frac{2\epsilon x}{(x^2+\epsilon^2)^2}=-\pi\delta^{\prime}(x). \label{delrep2}
\end{align}
Now
\begin{align}
\lim\limits_{\epsilon\rightarrow 0}\mbox{Im}\,\frac{1}{x+i\epsilon}=\frac{1}{2i}\lim\limits_{\epsilon\rightarrow 0}\left[\frac{1}{x+i\epsilon}-\frac{1}{x-i\epsilon}\right]=\frac{1}{2i}\lim\limits_{\epsilon\rightarrow 0}\frac{-2i\epsilon}{x^2+\epsilon^2}=-\lim\limits_{\epsilon\rightarrow 0}\frac{\epsilon}{x^2+\epsilon^2}=-\pi\delta(x),
\end{align}
and
\begin{align}
\lim\limits_{\epsilon\rightarrow 0}\mbox{Im}\,\frac{1}{(x+i\epsilon)^2}=\frac{1}{2i}\lim\limits_{\epsilon\rightarrow 0}\left[\frac{1}{(x+i\epsilon)^2}-\frac{1}{(x-i\epsilon)^2}\right]=\frac{1}{2i}\lim\limits_{\epsilon\rightarrow 0}\frac{-4i\epsilon x}{(x^2+\epsilon^2)^2}=-\lim\limits_{\epsilon\rightarrow 0}\frac{2\epsilon x}{(x^2+\epsilon^2)^2}=\pi\delta^{\prime}(x).
\end{align}
This proves equation \eqref{theo1} and \eqref{theo2}.
With these it is easy to get the spectral representation for the free part:
\begin{align}
\rho_0^{(1)}(k_0,|\vec{k}|) = \frac{1}{\pi}\mbox{Im}\,\frac{1}{2}\left(\frac{1}{k_0-\omega_k+i\epsilon}+\frac{1}{k_0+\omega_k+i\epsilon}\right) = -\frac{\delta(k_0+\omega_k)+\delta(k_0-\omega_k)}{2} \label{rho^1_0},
\end{align}
\begin{align}
\rho_0^{(0)}(k_0,|\vec{k}|) = \frac{1}{\pi}\mbox{Im}\,\frac{1}{2\omega_k}\left(\frac{1}{k_0-\omega_k+i\epsilon}-\frac{1}{k_0+\omega_k+i\epsilon}\right) = \frac{\delta(k_0+\omega_k)-\delta(k_0-\omega_k)}{2\omega_k} \label{rho^0_0}.
\end{align}
Now for the $1^{\mbox{st}}$ order part, we need to 
\begin{align}
	\frac{k_0}{(k^2_0-\omega^2_k)^2}=\frac{1}{4\omega_k}\frac{4k_0\omega_k}{(k_0+\omega_k)^2(k_0-\omega_k)^2}=\frac{1}{4\omega_k}\frac{(k_0+\omega_k)^2-(k_0-\omega_k)^2}{(k_0+\omega_k)^2(k_0-\omega_k)^2}=\frac{1}{4\omega_k}\left[\frac{1}{(k_0-\omega_k)^2}-\frac{1}{(k_0+\omega_k)^2}\right],
\end{align}
\begin{align}
\frac{1}{(k^2_0-\omega^2_k)^2}&=\frac{1}{4\omega^2_k}\left[\frac{1}{k_0-\omega_k}-\frac{1}{k_0+\omega_k}\right]^2=\frac{1}{4\omega^2_k}\left[\frac{1}{(k_0-\omega_k)^2}+\frac{1}{(k_0+\omega_k)^2}-\frac{2}{k^2_0-\omega^2_k}\right] \nonumber\\
&=\frac{1}{4\omega^2_k}\left[\frac{1}{(k_0-\omega_k)^2}+\frac{1}{(k_0+\omega_k)^2}-\frac{1}{\omega_k}\left(\frac{1}{k_0-\omega_k}-\frac{1}{k_0+\omega_k}\right)\right].
\end{align}
Thus
\begin{align}
\rho_1^{(1)}(k_0,|\vec{k}|)=\frac{1}{\pi}\mbox{Im}\frac{1}{4\omega_k}\left[\frac{1}{(k_0-\omega_k+i\epsilon)^2}-\frac{1}{(k_0+\omega_k+i\epsilon)^2}\right]=\frac{\delta^{\prime}(k_0-\omega_k)-\delta^{\prime}(k_0+\omega_k)}{4\omega_k} \label{rho^1_1}.
\end{align}
Also
\begin{align}
\rho_1^{(0)}(k_0,|\vec{k}|)&=\frac{1}{\pi}\mbox{Im}\,\frac{1}{4\omega^2_k}\left[\frac{1}{(k_0-\omega_k+i\epsilon)^2}+\frac{1}{(k_0+\omega_k+i\epsilon)^2}-\frac{1}{\omega_k}\left(\frac{1}{k_0-\omega_k+i\epsilon}-\frac{1}{k_0+\omega_k+i\epsilon}\right)\right] \nonumber\\
&=\frac{1}{4\omega^2_k}\left\lbrace \delta^{\prime}(k_0-\omega_k)+\delta^{\prime}(k_0+\omega_k)+\frac{1}{\omega_k}\left[\delta(k_0-\omega_k)-\delta(k_0+\omega_k)\right] \right\rbrace \label{rho^0_1}.
\end{align}
%

\end{document}